\documentclass[11pt]{article}
\pdfoutput=1
\usepackage{jheppub}
\usepackage{subfig}

\newcommand\hare[1]{\textcolor{red}{(hare: #1)}}

\title{Modified supersymmetric indices in AdS$_3$/CFT$_2$}

\date{}

\begin{document}
\author{ Arash Arabi Ardehali}
\author{and Hare Krishna}

\affiliation{C.N. Yang Institute for Theoretical Physics,\\ SUNY, Stony Brook, NY 11794, USA}

\emailAdd{a.a.ardehali@gmail.com}\emailAdd{harekrishna.harekrishna@stonybrook.edu}

\abstract{We consider the $\mathcal{N}=(2,2)$ AdS$_3$/CFT$_2$ dualities proposed by Eberhardt, where the bulk geometry is AdS$_3\times(S^3\times T^4)/\mathbb{Z}_k$, and the CFT is a deformation of the symmetric orbifold of the supersymmetric  sigma model $T^4/\mathbb{Z}_k$ (with $k=2,\ 3,\ 4,\ 6$). The elliptic genera of the two sides vanish due to fermionic zero modes, so for microstate counting applications one must consider modified supersymmetric indices. In an analysis similar to that of Maldacena, Moore, and Strominger for the standard $\mathcal{N}=(4,4)$ case of $T^4$, we study the appropriate helicity-trace index of the boundary CFTs. We encounter a strange phenomenon where a saddle-point analysis of our indices reproduces only a fraction (respectively $\frac{1}{2},\ \frac{2}{3},\ \frac{3}{4},\ \frac{5}{6}$) of the Bekenstein-Hawking entropy of the associated macroscopic black branes.}

\maketitle

\section{Introduction}

In \cite{Eberhardt:2017uup} Eberhardt introduced eight explicit instances of $\mathcal{N}=(2,2)$ AdS$_3$/CFT$_2$, where the bulk geometry is AdS$_3\times (S^3\times M_4)/G$, and the boundary CFT is (a marginal deformation of) the symmetric orbifold $\mathrm{sym}^N(M_4/G)$. The group $G$ is finite and freely acting in all eight cases. In one case $M_4=K3$ and $G=\mathbb{Z}_2$, such that $M_4/G=K3/\mathbb{Z}_2$ is the Enriques surface denoted $\mathrm{ES}$. In the remaining cases $M_4=T^4$, and each of the seven choices for $G$ corresponds to a complex hyperelliptic surface $\mathrm{HS}=T^4/G$ as reviewed in Appendix~\ref{app:hodge}. Setting $G=1$ recovers of course the standard $\mathcal{N}=(4,4)$ dualities \cite{Maldacena:1997re}. See \cite{Aharony:1999ti,David:2002wn} for reviews of the $\mathcal{N}=(4,4)$ cases.

We are interested in these dualities because they have \emph{smooth} supergravity geometries in the bulk, and \emph{explicit} CFT descriptions on the boundary. Eberhardt's paper grew out of the earlier work \cite{Datta:2017ert} where $\mathcal{N}=(2,2)$ examples with $M=T^4$ and $G$ a dihedral group were considered but the bulk geometries were not generally smooth. See also \cite{Eberhardt:2018ouy,Gaberdiel:2019wjw} for dualities between a tensionless limit of the bulk theories and undeformed symmetric orbifolds on the boundary. Other examples of $\mathcal{N}=(2,2)$ AdS$_3$/CFT$_2$ where the boundary theory is only implicit as the IR fixed point of an RG flow across dimensions can be found in \cite{Couzens:2017nnr,Couzens:2021tnv}.
For studies of more general structures in $\mathcal{N}=(2,2)$ holography involving symmetric orbifold CFTs see \cite{Benjamin:2015hsa,Belin:2019rba,Belin:2020nmp,Benjamin:2022jin} and references therein.\\

Aspects of these dualities have passed successful checks in \cite{Eberhardt:2017uup,Datta:2017ert,ArabiArdehali:2021iwe}:
\begin{itemize}
    \item the bulk BPS Kaluza-Klein (KK) spectra were matched in \cite{Eberhardt:2017uup} (in the $T^4/\mathbb{Z}_2$ case even earlier, see \cite{Datta:2017ert}) with those of the corresponding boundary operators;
    \item in \cite{ArabiArdehali:2021iwe} associated brane configurations were proposed whose low-energy sigma models reproduced the bulk central charges, at the leading ($\mathcal{O}(1/G_N)$) as well as the subleading ($\mathcal{O}(G_N^0)$) order \cite{Beccaria:2014qea,ArabiArdehali:2018mil} in the bulk Newton's constant $G_N$;
    \item in \cite{ArabiArdehali:2021iwe} the nontrivial boundary elliptic genus in the $\mathrm{ES}$ case was used for microstate counting of the bulk black holes, and the Bekenstein-Hawking entropy ($\mathcal{O}(1/G_N)$) as well as the one-loop logarithmic correction to it ($\mathcal{O}(\log G_N)$) were matched.
\end{itemize}

In the HS cases, the elliptic genera vanish due to fermionic zero modes. So for black hole microstate counting one must appeal instead to variants of the elliptic genus. In this work we initiate the analysis needed for solving this problem.

The issue with fermionic zero modes and the need for modified indices has in fact already been encountered in the standard $\mathcal{N}=(4,4)$ case where $M_4=T^4.$ In that case there are two fermionic zero modes---one for each $T^2$---and Maldacena-Moore-Strominger (MMS) \cite{Maldacena:1999bp} proposed considering a modified index containing two helicity insertions (see \eqref{eq:MMSindex} for the precise expression). The idea is that each helicity insertion soaks up a fermionic zero mode, so while the usual index is killed by the zero modes, this ``second helicity-trace index'' survives. MMS moreover showed that long representations of the right-handed $\mathcal{N}=4$ algebra do not contribute to the modified index, so it ought to be protected against quantum corrections and suitable for black hole microstate counting.

Our method is essentially that of MMS adapted to the $\mathcal{N}=(2,2)$ context. The target spaces $\mathrm{HS}=T^4/G$ of our interest have only one fermionic zero mode, so for us a single helicity insertion in the index would do. We thus consider \emph{the first helicity-trace index}, which we demonstrate ought to be protected with $\mathcal{N}=(2,2)$ superconformal symmetry.

We focus here on the four cases where $G$ is simply $\mathbb{Z}_k$ (with $k=2,3,4,6),$ leaving the three remaining cases for future work. Elementary CFT orbifold techniques allow us to adapt the MMS analysis to these $T^4/\mathbb{Z}_k$ cases. To our surprise, we find that a saddle-point analysis of our boundary indices reproduces only a fraction (respectively $\frac{1}{2},\ \frac{2}{3},\ \frac{3}{4},\ \frac{5}{6}$) of the bulk Bekenstein-Hawking entropies.

The saddle-point analysis also yields a logarithmic correction to the Bekenstein-Hawking entropy, which turns out to match the macroscopic expectation from a one-loop calculation on the near-horizon geometry \cite{Sen:2012cj} presented in Section~\ref{subsec:bulkEntropies}.

In the $k=2$ case, it turns out that our helicity-trace index is related to the weak Jacobi form $\phi_{-1,2}$ of weight $-1$ and index $2$. This makes available an alternative approach to microstate counting of the bulk black branes via Rademacher expansion of $\phi_{-1,2}$. This way we corroborate the results of our saddle-point analysis for $k=2$.

We also provide numerical evidence for validity of our saddle-point results for all $k$. Our conclusion is thus that Bose-Fermi cancellations prevent the helicity-trace index to capture the full Bekenstein-Hawking entropy in these dualities. The standard argument for supersymmetric indices encoding the full entropy \cite{Dabholkar:2010rm} relies on an $\mathrm{SU}(2)_R$ near-horizon symmetry, and seems to be evaded in our $\mathcal{N}=(2,2)$ cases due to the $\mathrm{SU}(2)_R$ being broken to $U(1)_R$.\\

Here is a brief outline of what follows. We begin in Section~\ref{sec:helicityTrace} by demonstrating that the non-trivial SUSY index in the $(2,2)$ context of our interest is the 1st helicity-trace index, denoted $\mathcal{E}_1$. We proceed to explain how it is evaluated for Eberhardt's symmetric orbifold CFTs $\mathrm{sym}^N (T^4/\mathbb{Z}_k)$. The calculation starts with finding the---$(-1)^F$ weighted---partition function $Z$ of the seed SUSY sigma models $T^4/\mathbb{Z}_k=:\mathrm{HS}_k$. This is done in Section~\ref{subsec:seed} using elementary orbifold (S)CFT techniques found in \cite{DiFrancesco:1997nk} (the ``Yellow Book'') and~\cite{Hori:2003ic} (the Mirror Symmetry book). Going from the seed partition function $Z$ to the symmetric orbifold index $\mathcal{E}_1$ is achieved in Section~\ref{subsec:DMVV} via the DMVV formula \cite{Dijkgraaf:1996xw}, whose content is summarized neatly---see \eqref{eq:E1cHat1}---in terms of the Fourier coefficients $\hat{c}_1$ of a counting function $\mathcal{H}_1$ introduced in Section~\ref{subsec:countingFn}. The asymptotic degeneracies encoded in the modified indices $\mathcal{E}_1$ are thus found from the asymptotic content of the counting functions $\mathcal{H}_1[\mathrm{HS}_k],$ which we study in Section~\ref{sec:asy degeneracy} and compare with macroscopic expectations. We find in particular that in the $\mathrm{HS}_k$ case log of the asymptotic degeneracies encoded in the modified index $\mathcal{E}_1$ accounts only for a fraction $1-\frac{1}{k}$ of the Bekenstein-Hawking entropy of the bulk black branes, while it successfully reproduces the logarithmic correction to it. Section~\ref{sec:Discussion} summarizes our understanding of the achievements and challenges in Eberhardt's $(2,2)$ dualities. The four appendices contain technical details suppressed in the main text.

\section{Helicity-trace index for $\mathcal{N}=(2,2)$ CFTs}\label{sec:helicityTrace}

In their study of the $\mathcal{N}=(4,4)$ CFT with target space $\mathrm{sym}^N(T^4)$, MMS introduced the second helicity-trace index defined as \cite{Maldacena:1999bp}
\begin{equation}
    \mathcal{E}_2:= \mathrm{Tr}_{RR}\Big[(-1)^{2 J_0^3-2 \bar{J}_0^3} (2 \bar{J}_0^3)^2 q^{L_0} \bar{q}^{\bar{L}_0} y^{2 J_0^3}\Big]\label{eq:MMSindex}.
\end{equation}
Here $J^3_0$ and $\bar J^3_0$ are the Cartans of the left- and right-handed $\mathrm{SU}(2)_R$ current zero modes, while $L_0$ and $\bar L_0$ are the left- and right-handed Virasoro zero modes, and the trace is over the Ramond-Ramond (RR) sector.

They demonstrated that long representations of the (right-handed) $\mathcal{N}=4$ algebra do not contribute to $\mathcal{E}_2$. Therefore $\mathcal{E}_2$ is expected to be protected against quantum corrections arising from marginal deformation of the symmetric orbifold to the strong coupling regime dual to string theory on AdS$_3\times S^3\times T^4$. They moreover showed that, unlike the elliptic genus, the second helicity-trace of the $\mathcal{N}=(4,4)$ $\mathrm{sym}^N(T^4)$ CFT is nonzero, because the two insertions of $2 \bar{J}_0^3$ inside the trace soak up the two fermionic zero modes of $T^4$. In this section, we are going to adapt this analysis to the $\mathcal{N}=(2,2)$ cases with target space $\mathrm{sym}^N(\mathrm{HS})$.

Of the two fermionic zero modes of $T^4,$ one survives the action of $G$. To see this, we write $T^4=T^2\times T^2$, and note that in all $\mathrm{HS}$ cases $G$ shifts the first $T^2$ (leaving its fermionic zero mode intact) while acting as in Appendix~\ref{app:hodge} non-trivially on the second $T^2$ (killing its fermionic zero mode). We will hence need only one helicity insertion inside the trace: 
\begin{equation}
    \mathcal{E}_1:= \mathrm{Tr}_{RR}\Big[(-1)^{2 J_0-2 \bar{J_0}} (2 \bar{J_0}) q^{L_0} \bar{q}^{\bar{L}_0} y^{2 J_0}\Big],
\end{equation}
with $J_0$ and $\bar J_0$ the left- and right-handed $U(1)_R$ current zero modes. This \emph{first helicity-trace index} is zero for the $(4,4)$ case of $T^4$ (because it does not soak up one of its fermionic zero modes), but will be non-zero for the $(2,2)$ cases of our interest. It is moreover protected against quantum corrections, as we now demonstrate.

Let us take a closer look at the right-handed supersymmetry algebra. This is the part of the SUSY algebra responsible for the cohomological structure with respect to which we want to show $\mathcal{E}_1$ is an index. In the context of our interest, where there is a fermionic zero mode, the CFT target space has a $U(1)^2$ isometry which descends to a current algebra enhancing the usual $\mathcal{N}=2$ symmetry. The zero-mode part of this enhanced algebra takes the form (compare with Eq.~(3.9) of \cite{Maldacena:1999bp}):
 \begin{eqnarray}
 \label{zero_mode_susy}
    &&\lbrace \bar{G}_0^+,\bar{G}_0^-\rbrace= 2 \bar{L}_0, \quad  \lbrace \bar{Q}_0^+,\bar{Q}_0^-\rbrace= 1,\nonumber\\
   &&  [\bar{J}_0,\bar{G}_0^+]=\frac{1}{2} \bar{G}_0^+,\quad \ [\bar{J}_0,\bar{G}_0^-]=-\frac{1}{2} \bar{G}_0^-,\nonumber\\
    &&  [\bar{J}_0,\bar{Q}_0^+]=\frac{1}{2} \bar{Q}_0^+,\quad \ [\bar{J}_0,\bar{Q}_0^-]=-\frac{1}{2} \bar{Q}_0^-.
 \end{eqnarray}
Here $G^{\pm}$ are the fermionic superpartners of the right-handed stress tensor, while $\bar{J}$ is the right-handed $U(1)_R$ current, and $\bar{Q}^{\pm}$
the fermionic superpartners of the right-handed $U(1)^2$ current. The commutators of $\bar{G}$s with $\bar{Q}$s vanish if we just consider states that are neutral under $U(1)^2$. We now demonstrate that given the zero mode algebra~\eqref{zero_mode_susy}, long representations do not contribute to $\mathcal{E}_1$.

For a long representation, we have $\bar{L}_0>0$. The $G$s and $Q$s are fermionic in nature, so they will behave as fermionic raising and lowering operators. The algebra~\eqref{zero_mode_susy} has two fermionic creation operators $b_i^{\dagger}$, corresponding to $\bar{G}^+_0$ and $\bar{Q}^{+}_0$, which have $\bar{J}_0=\frac{1}{2}$. The corresponding lowering operators $b_i$ have $\bar{J}_0=-\frac{1}{2}$. Let us start with the state $|0,j\rangle$ annihilated by the lowering operators and obeying $\bar{J}_0|0,\bar j\rangle=\bar j|0,\bar j\rangle$. Acting on it with creation operators we get two states with $\bar{J}_0=\bar j+\frac{1}{2}$, and one state with $\bar{J}_0=\bar j+1$. The fermion number $F:=2J_0-2\bar{J}_0$ of these states alternates. We can hence easily check that the long representations do not contribute to the first helicity-trace index
\begin{equation}
\mathrm{Tr}^{}_{{\text{long}}}(-1)^{-2\bar J_0}\, \bar J_0\propto \bar j-2(\bar j+\frac{1}{2})+ \bar j+1 =0.\label{eq:longSum0}
\end{equation}
Thus $\mathcal{E}_1$ is a protected index in the $\mathcal{N}=(2,2)$ theories of our interest.\footnote{Actually, this argument does not yet guarantee protectedness of the $L_0,J_0$ quantum numbers of long$\times$short representations. For that we appeal to the assumptions $i$) $L_0-\bar L_0\in\mathbb{Z},$ and $ii$) quantization of $J_0$ charge.}

We compute such helicity-trace indices by taking $\bar{y}$ derivatives of the partition function\footnote{As in \cite{Maldacena:1999bp} we use a notation where $L_0,\bar{L}_0$ are zero on RR ground states. The reader familiar with the alternative notation might want to replace $L_0,\bar{L}_0$ with $L_0-c/24,\bar{L}_0-c/24$ in the following formula.}
\begin{equation}
Z(q,\bar{q},y,\bar{y})=\mathrm{Tr}_{RR}(-1)^{2J_0-2\bar{J}_0}q^{L_0}\bar{q}^{\bar{L}_0}y^{2J_0}\bar{y}^{2\bar{J}_0},\label{eq:Zdef}
\end{equation}
and then setting $\bar{y}=1$.

We would like to compute the helicity-trace index of a symmetric orbifold CFT with target space $\mathrm{sym}^N M_4$. For that we need the partition function $Z[\mathrm{sym}^N  M_4]$, which can be extracted from the DMVV formula \cite{Dijkgraaf:1996xw}
\begin{eqnarray}
    \mathcal{Z}[M_4]=\sum_{N=0}^\infty p^N Z[\mathrm{sym}^N  M_4]=\prod_{i=1}^{\infty} \prod_{\Delta,\bar{\Delta},\ell,\bar{\ell}}^{'} \frac{1}{(1-p^i q^{{\Delta}/{i}} \bar{q}^{{\bar{\Delta}}/{i}} y^{\ell} \bar{y}^{\bar{\ell}})^{c(\Delta,\bar{\Delta},\ell,\bar{\ell})}}.\label{eq:DMVVog}
\end{eqnarray}
Here the prime indicates that $\Delta,\bar{\Delta}$ are restricted so that $\frac{\Delta-\bar\Delta}{i}\in\mathbb{Z}$, while $c(\Delta,\bar{\Delta},\ell,\bar{\ell})$ denotes the coefficients of the partition function
\begin{eqnarray}
Z[M_4]=\sum_{\Delta,\bar{\Delta},\ell,\bar{\ell}} c(\Delta,\bar{\Delta},\ell,\bar{\ell}) q^{\Delta}\bar{q}^{\bar{\Delta}}y^{\ell} \bar{y}^{\bar{\ell}},\label{eq:cDef}
\end{eqnarray}
of the seed CFT with target space $M_4.$

\subsection{Partition function of the seed theory}\label{subsec:seed}

The seed CFTs that we consider in this work have target spaces of the form $M_4=T^4/G$, with $G=\mathbb{Z}_k$. Writing $T^4=T^2\times T^2$, the group $G$ shifts the first $T^2$ and rotates the second $T^2$. Since the order-$k$ rotation has to be an automorphism of the lattice of the torus, the complex structure of the second $T^2$ must take specific values depending on $k$ as in Appendix~\ref{app:hodge} (except in the $\mathbb{Z}_2$ case where it can be arbitrary). This is one of the reasons we have only $k=2,3,4,6$.

Let $g$ be the fundamental generator of $G.$ The partition function of the $T^4/G$ CFT is given by the standard orbifold formula
\begin{equation}
    Z[T^4/\mathbb{Z}_k]=\frac{1}{k}\big(\sum_{j,t=0}^{k-1} Z^{g^t\text{-twisted}}_{\text{w/\
}g^j\text{-insertion}}[T^4]\big),\label{eq:orbGeneralFormula}
\end{equation}
with $t$ labeling the twisted sectors.

\subsubsection*{Low-temperature limit}

In the next section we will study the Cardy-like ($q,\bar{q}\to1$, or ``high-temperature'') limit of the $\mathrm{sym}^N M_4$ index to extract microstate degeneracies of the associated macroscopic black branes. Here, we point out a simple consistency check for the $q,\bar{q}\to0$ (or ``low-temperature'') limit of the $M_4$ partition function.

This limit has to capture the supersymmetric RR sector ground states contributing to $Z[M_4].$ These ground states are in one-to-one correspondence with the harmonic $(\mathrm{p},\mathrm{q})$ forms on the  target space $M_4$, and since $M_4$ is two-complex-dimensional, they carry $R$-charges\footnote{We are following the conventions of \cite{Hori:2003ic}; see Eqs.~(13.72)--(13.73) therein. Note that $q^{\mathrm{there}}_R=2J_0$ and $q^{\mathrm{there}}_L=2\bar{J}_0$; these relations are obtained comparing the $R$-charges of fermions: $2J_0=1$ for $\psi,$ and $2\bar{J}_0=1$ for $\tilde{\psi}$ (see Eqs.~(11.162)--(11.163) in \cite{Hori:2003ic}).}
\begin{equation}
    2J_0=\mathrm{q}-\frac{\mathrm{dim}_{\mathbb{C}}M}{2}=\mathrm{q}-1,\qquad 2\bar{J}_0=\mathrm{p}-\frac{\mathrm{dim}_{\mathbb{C}}M}{2}=\mathrm{p}-1.
\end{equation} 

Hence the coefficient of $(-1)^{2J_0-2\bar{J}_0}y^{2J_0}\bar{y}^{2\bar{J}_0}$ in $\lim_{q,\bar{q}\to0}Z[M_4]$ should reproduce $h_{2\bar{J}_0+1,2J_0+1}$ in the Hodge diamond of $M_4$. Said differently, for each harmonic $(\mathrm{p},\mathrm{q})$ form on $M_4$ there should be a term $(-1)^{\mathrm{q}-\mathrm{p}} y^{\mathrm{q}-1}\bar{y}^{\mathrm{p}-1}$ in the low-temperature limit of  $Z[M_4]$.

\subsubsection*{Examples}

We start with $T^4/\mathbb{Z}_2$, which we denote as $\mathrm{HS}_2.$ The formula \eqref{eq:orbGeneralFormula} can be worked out in detail and as we demonstrate in Appendix~\ref{app:seed} it yields
\begin{equation}
\begin{split}
Z[\mathrm{HS}_2]&=\frac{1}{2}\Theta^{T^4}\left|\frac{\theta_1(z,\tau)}{\eta^3}\right|^4
+2\left|\frac{\theta_2(z,\tau)}{\theta_2(\tau)}\right|^2\cdot
\Theta^{T^2}_{\text{w/\ }G\text{-insertion}}\cdot
\left|\frac{\theta_1(z,\tau)}{\eta^3}\right|^2\\
&\ \ +2\left|\frac{\theta_4(z,\tau)}{\theta_4(\tau)}\right|^2\cdot
\Theta^{T^2}_{\text{w/\ }1/2-\text{shifted lattice}}\cdot
\left|\frac{\theta_1(z,\tau)}{\eta^3}\right|^2\\
&\ \ +2\left|\frac{\theta_3(z,\tau)}{\theta_3(\tau)}\right|^2\cdot
\Theta^{T^2}_{\text{w/\ }G\text{-insertion, }1/2-\text{shifted
lattice}}\cdot \left|\frac{\theta_1(z,\tau)}{\eta^3}\right|^2.\label{eq:Zhs2}
\end{split}
\end{equation}

We are using the convention $\theta_i(\tau):=\theta_i(0,\tau)$. The
$\Theta$ functions involve sums over lattices of the topological
sectors of the compact scalar fields---see Appendix~\ref{app:seed} for more details. As in \cite{Maldacena:1999bp},
we restrict to the origin of the lattice (otherwise there would be non-trivial dependence on $\bar{q}$ in the modified index, in contrast to it being a BPS index with respect to right-handed supersymmetry). This amounts to suppressing the
$\Theta$ functions in the first line, and completely neglecting
the last two lines of the above equation. We end up with
\begin{equation}
Z[\mathrm{HS}_2]_{\text{singlet}}=\frac{1}{2}\left|\frac{\theta_1(z,\tau)}{\eta^3}\right|^4
+2\left|\frac{\theta_2(z,\tau)}{\theta_2(\tau)}\right|^2\cdot
\left|\frac{\theta_1(z,\tau)}{\eta^3}\right|^2.\label{eq:HSpfZ2}
\end{equation}
As a consistency check, we note that in the low-temperature $q,\bar{q}\to 0$ limit,
\begin{equation}
Z[\mathrm{HS}_2]_{\text{singlet}}\to
\frac{1}{2}(y^{1/2}-y^{-1/2})^2(\bar{y}^{1/2}-\bar{y}^{-1/2})^2+\frac{1}{2}(y-y^{-1})(\bar{y}-\bar{y}^{-1}),\label{eq:HSpfZ2Test}
\end{equation}
with the coefficient of $(-1)^{2J_0-2\bar{J}_0}y^{2J_0}\bar{y}^{2\bar{J}_0}$ correctly reproducing $h_{2\bar{J}_0+1,2{J}_0+1}$ in the Hodge diamond of $\mathrm{HS}_2$ as presented in Appendix~\ref{app:hodge}.\\

For the $G=\mathbb{Z}_{3,4,6}$ cases also we relegate the derivations to Appendix~\ref{app:seed} and only present the final results here:
\begin{equation}
\begin{split}
Z[\mathrm{HS}_3]&=\frac{1}{3}\Theta^{T^4}\left|\frac{\theta_1(z,\tau)}{\eta^3}\right|^4
+\sum_{j=1,2}\left|\frac{\theta_1(z+\frac{j}{3},\tau)}{\theta_1(\frac{j}{3},\tau)}\right|^2\cdot
\Theta^{T^2}_{\text{w/\ }g^j\text{-insertion}}\cdot
\left|\frac{\theta_1(z,\tau)}{\eta^3}\right|^2\\
&+\sum_{t=1,2}\sum_{j=0}^{2}(y\bar y)^{1-\frac{t}{3}}\left|\frac{\theta_1(z+\frac{t}{3}\tau+\frac{j}{3},\tau)}{\theta_1(\frac{t}{3}\tau+\frac{j}{3},\tau)}\right|^2\cdot\Theta^{T^2}_{\text{w/ }g^j\text{-insertion,  }t/3\text{-shifted
lattice}}\cdot
\left|\frac{\theta_1(z,\tau)}{\eta^3}\right|^2,
\label{eq:Zhs3}
\end{split}
\end{equation}

\begin{equation}
\begin{split}
&Z[\mathrm{HS}_4]=\frac{1}{4}\Bigg(\Theta^{T^4}\left|\frac{\theta_1(z,\tau)}{\eta^3}\right|^4+\sum_{j=1}^3 4\sin^2[\frac{\pi j}{4}]\left|\frac{\theta_1(z+\frac{j}{4},\tau)}{\theta_1(\frac{j}{4},\tau)}\right|^2\cdot
\Theta^{T^2}_{\text{w/\ }g^j\text{-insertion}}\cdot
\left|\frac{\theta_1(z,\tau)}{\eta^3}\right|^2\\
&+\sum_{t=1,3}\sum_{j=0}^{3}2(y\bar y)^{1-\frac{t}{4}}\left|\frac{\theta_1(z+\frac{t}{4}\tau+\frac{j}{4},\tau)}{\theta_1(\frac{t}{4}\tau+\frac{j}{4},\tau)}\right|^2\cdot\Theta^{T^2}_{\text{w/ }g^j\text{-insertion, }t/4-\text{shifted
lattice}}\cdot
\left|\frac{\theta_1(z,\tau)}{\eta^3}\right|^2\\
&+\sum_{\substack{j=0\\
t=2}}^{3}3(y\bar y)^{1-\frac{t}{4}}\left|\frac{\theta_1(z+\frac{t}{4}\tau+\frac{j}{4},\tau)}{\theta_1(\frac{t}{4}\tau+\frac{j}{4},\tau)}\right|^2\cdot\Theta^{T^2}_{\text{w/ }g^j\text{-insertion, }t/4-\text{shifted
lattice}}\cdot
\left|\frac{\theta_1(z,\tau)}{\eta^3}\right|^2\Bigg),
\label{eq:Zhs4}
\end{split}
\end{equation}

\begin{equation}
\begin{split}
Z[\mathrm{HS}_6]&=\frac{1}{6}\Bigg(\Theta^{T^4}\left|\frac{\theta_1(z,\tau)}{\eta^3}\right|^4
+\sum_{j=1}^5 4 \sin^2[\frac{\pi j}{6}]\left|\frac{\theta_1(z+\frac{j}{6},\tau)}{\theta_1(\frac{j}{6},\tau)}\right|^2\cdot
\Theta^{T^2}_{\text{w/\ }g^j\text{-insertion}}\cdot
\left|\frac{\theta_1(z,\tau)}{\eta^3}\right|^2\\
&+\sum_{t=1,5}\sum_{j=0}^{5}(y\bar y)^{1-\frac{t}{6}}\left|\frac{\theta_1(z+\frac{t}{6}\tau+\frac{j}{6},\tau)}{\theta_1(\frac{t}{6}\tau+\frac{j}{6},\tau)}\right|^2\cdot\Theta^{T^2}_{\text{w/ }g^j\text{-insertion, }t/6-\text{shifted
lattice}}\cdot
\left|\frac{\theta_1(z,\tau)}{\eta^3}\right|^2\\
&+\sum_{t=2}^4\sum_{j=0
}^{5}2(y\bar y)^{1-\frac{t}{6}}\left|\frac{\theta_1(z+\frac{t}{6}\tau+\frac{j}{6},\tau)}{\theta_1(\frac{t}{6}\tau+\frac{j}{6},\tau)}\right|^2\cdot\Theta^{T^2}_{\text{w/ }g^j\text{-insertion, }t/6-\text{shifted
lattice}}\cdot
\left|\frac{\theta_1(z,\tau)}{\eta^3}\right|^2\Bigg).
\label{eq:Zhs6}
\end{split}
\end{equation}

We have checked that their low-temperature limit coincides with \eqref{eq:HSpfZ2Test}, and therefore reproduces the correct $\mathrm{HS}$ Hodge diamond as in Appendix~\ref{app:hodge}.

\subsection{Index of the symmetric orbifold CFT}\label{subsec:DMVV}

Taking the derivative of $\mathcal{Z}[M_4]$ with respect to $\bar{y}$ and setting $\bar{y}=1$ gives the generating function of the modified index. From Eq.~\eqref{eq:DMVVog} we get for the derivative
\begin{equation}
\begin{split}
\partial_{\bar{y}}  \mathcal{Z}[M_4]\big|_{\bar{y}=1}&=\partial_{\bar{y}} \left(\prod_{i=1}^{\infty} \prod_{\Delta,\bar{\Delta},\ell,\bar{\ell}}^{'} \frac{1}{(1-p^i q^{\Delta/i} \bar{q}^{\bar{\Delta}/i} y^{\ell} \bar{y}^{\bar{\ell}})^{c(\Delta,\bar{\Delta},\ell,\bar{\ell})}}\right)_{\bar{y}=1}\\
&= \sum_{i,\Delta,\bar{\Delta},\ell,\bar{\ell}}^{'}\frac{\bar{\ell}\, c(\Delta,\bar{\Delta},\ell,\bar{\ell})\, p^i q^{\Delta/i} \bar{q}^{\bar{\Delta}/i} y^\ell}{1-p^i q^{\Delta/i} \bar{q}^{\bar{\Delta}/i} y^\ell}\mathcal{Z}[M_4]\bigg|_{\bar{y}=1}\\
&=\sum_{i \in \mathbb{N}, \Delta\geq 0,\ell} \frac{\hat{c}_1(i\Delta,\ell) p^i q^\Delta y^{\ell}}{1- p^i q^\Delta y^{\ell}},
\label{eq:partialYbarCurlyZ}
\end{split}
\end{equation}
Above we have used $\mathcal{Z}[M_4]\big|_{\bar{y}=1}=1,$ following from the fact that the elliptic genus of $\mathrm{sym}^N M_4$ vanishes for $N>0$ in the present context. We have also defined
\begin{eqnarray}
    \hat{c}_1 (\Delta,\ell):=\sum_{\bar{\ell}}\bar{\ell} c(\Delta,0,\ell,\bar{\ell}),\label{eq:cHat1Def}
\end{eqnarray}
and used
\begin{equation}
\sum_{\bar{\ell}}\bar{\ell} c(\Delta,\bar{\Delta},\ell,\bar{\ell})=0, \quad \text{for} \quad \bar{\Delta}>0.
\end{equation}
The latter identity can be checked for $M_4=\mathrm{HS}_{2,3,4,6}$ at low values of $\Delta$ for various cases by working out the expansion of $Z[M_4]$ as in \eqref{eq:cDef}, and then explicitly evaluating the sum. More generally, it follows from~\eqref{eq:longSum0}, noting that the creation operators $\bar{G}^+_0$ and $\bar{Q}^{+}_0$ commute with $L_0,\bar L_0, J_0$.

Now we can expand out \eqref{eq:partialYbarCurlyZ} as
\begin{eqnarray}
 \partial_{\bar{y}}  \mathcal{Z}[M_4]\big|_{\bar{y}=1}=\sum_{N=1}^\infty p^N \mathcal{E}_1[\mathrm{sym}^N M_4]=\sum_{s,i \in \mathbb{N}, \Delta\geq 0,\ell} \hat{c}_1(i\Delta,\ell)( p^i q^\Delta y^{\ell})^s .\label{eq:modifiedDMVV}
\end{eqnarray}
Let us restrict attention to a term on the RHS with the power of $q$ a prime number $n$, and the power of $y$ another prime number $j$ (limiting to prime numbers is only for simplicity; we will rectify this limitation in Appendix~\ref{app:fine-grained} where we allow $s$ to take any natural number and observe interesting fine-grained number-theoretic structures).  Then we see that it is the $s = 1,\, i =
N,\, \Delta = n,\, \ell = j $ term that encodes the degeneracy in the modified index. We summarize this as
\begin{equation}
    \text{degeneracy of the states with $\Delta=n$, $\ell=j$ encoded in }\mathcal{E}_1[\mathrm{sym}^N M_4]\ \longrightarrow\ \hat{c}_1(N n,j).\label{eq:E1cHat1}
\end{equation} 

Our desired $\hat{c}_1$ coefficients can be computed via their definition \eqref{eq:cHat1Def}, using the $c$ coefficients of the non-holomorphic partition function $Z[M_4]$ as in \eqref{eq:cDef}. However, it turns out that the non-holomorphic $Z[M_4]$ contains too much irrelevant information for our purposes, and $\hat{c}_1(\Delta,\ell)$ is actually encoded as the coefficient of $q^\Delta y^\ell$ in the expansion of a certain \emph{holomorphic} two-variable function. The following simple calculation demonstrates this.

\subsection{The holomorphic counting function}\label{subsec:countingFn}

The non-holomorphic seed partition functions in our examples are (in the topologically trivial sector) of the form
\begin{equation}
    Z[\mathrm{HS}]=\sum_j B_j(q,y) \overline{B_j(q,y)}.
\end{equation}
Denote the coefficients of $B_j$ by $c_j^\text{hol}$:
\begin{equation}
B_j(q,y)=\sum_{\Delta,\ell}c^\text{hol}_j(\Delta,\ell)q^\Delta
y^\ell.\label{eq:cHolDef}
\end{equation}
Then the formula \eqref{eq:cHat1Def} for $\hat{c}_1$ can be evaluated as
\begin{equation}
\begin{split}
\hat{c}_1(\Delta,\ell)&=\sum_{\bar{\ell}}\bar{\ell}\big[\sum_j\overline{B_j}\cdot B_j\big]_{q^\Delta
\bar{q}^{0} y^\ell \bar{y}^{\bar{\ell}}}=\sum_j\big(\sum_{\bar{\ell}}\bar{\ell}\big[\,\overline{B_j}\,\big]_{\bar{q}^{0} \bar{y}^{\bar{\ell}}}\big)\left[B_j\right]_{q^\Delta
y^\ell }\\
&=\sum_j\big(\sum_{\bar{\ell}}\bar{\ell}c^\text{hol}_j(0,\bar{\ell})\big)\left[B_j\right]_{q^\Delta
y^\ell }=\left[\sum_j b_j\,  B_j\right]_{q^\Delta
y^\ell }.\label{eq:cHat1VScHolJ}
\end{split}
\end{equation}
The symbol $[f]_{x^k}$ above denotes the coefficient of $x^k$ in $f$. We have also defined
\begin{equation}
    b_j:=\sum_{\bar{\ell}}\bar{\ell}c^\text{hol}_j(0,\bar{\ell}).\label{eq:bjDef}
\end{equation} 

Eq.~\eqref{eq:cHat1VScHolJ} means  that $\hat{c}_1(\Delta,\ell)$ is encoded as the coefficient of $q^\Delta y^\ell$ in the series expansion of the holomorphic function
\begin{equation}
    \mathcal{H}_1(q,y):=\sum_j b_j\cdot B_j(q,y).
\end{equation}
These holomorphic functions are the objects of main interest in application to black hole microstate counting. Note that they are nothing but the 1st helicity-trace index of the seed CFT.

\subsubsection*{Examples}

For the $\mathrm{HS}_2$ case, we see from \eqref{eq:HSpfZ2} that
\begin{equation}
    Z[\mathrm{HS}_2]=\sum_{j=0,1}B_j\overline{B_j},
\end{equation}
with
\begin{equation}
    B_0=\frac{1}{\sqrt{2}}\left(\frac{\theta_1(z,\tau)}{\eta^3(\tau)}\right)^2,\qquad B_1=\sqrt{2}\, \frac{\theta_2(z,\tau)}{\theta_2(\tau)}\frac{\theta_1(z,\tau)}{\eta^3(\tau)}.
\end{equation}
Explicit evaluation of \eqref{eq:bjDef} shows\footnote{The coefficient $b_0$ is always zero in our examples, because it is proportional to the quantity $\sum_{\bar{\ell}}\bar{\ell}c_j^\text{hol}(0,\bar{\ell})$ computed for $T^4$, which is zero since $T^4$ needs two insertions of $\bar{\ell}$ to soak up its fermionic zero modes.} $b_0=0$ and $b_1=-\sqrt{2}i$ in this case. Hence
\begin{equation}
\boxed{\mathcal{H}_1[\mathrm{HS}_2]=-2i\,\frac{\theta_2(z,\tau)}{\theta_2(\tau)}\frac{\theta_1(z,\tau)}{\eta^3(\tau)}=y^{-1}-y+\mathcal{O}(q^2).}\label{eq:H1Z2result}
\end{equation}

For the $\mathrm{HS}_3$ case we have
\begin{eqnarray}
    Z[\mathrm{HS}_3]=\sum_{j=0,1,2}B_j \overline{B_j},
\end{eqnarray}
with
\begin{equation}
    B_0=\frac{1}{\sqrt{3}}\left(\frac{\theta_1(z,\tau)}{\eta^3(\tau)}\right)^2,\qquad B_1=\frac{\theta_1(z+\frac{1}{3},\tau)}{\theta_1(\frac{1}{3},\tau)}\frac{\theta_1(z,\tau)}{\eta^3},\quad B_2=\frac{\theta_1(z+\frac{2}{3},\tau)}{\theta_1(\frac{2}{3},\tau)}\frac{\theta_1(z,\tau)}{\eta^3}.
\end{equation}
Explicit evaluation of \eqref{eq:bjDef} gives $b_0=0,$  $b_1=- i,$  and $b_2= - i$. Hence
\begin{eqnarray}
      \boxed{\mathcal{H}_1[\mathrm{HS}_3]=-i  \Big(\frac{\theta_1(z+\frac{1}{3},\tau)}{\theta_1(\frac{1}{3},\tau)}\frac{\theta_1(z,\tau)}{\eta^3}+\frac{\theta_1(z+\frac{2}{3},\tau)}{\theta_1(\frac{2}{3},\tau)}\frac{\theta_1(z,\tau)}{\eta^3}\Big)=y^{-1}-y+\mathcal{O}(q).}\label{eq:H1Z3result}
\end{eqnarray}

For $\mathrm{HS}_4$ we have
\begin{eqnarray}
    Z[\mathrm{HS}_4]=\sum_{j=0,1,2,3} B_j \overline{B_j},
\end{eqnarray}
with 
\begin{eqnarray}
  B_0= \frac{1}{\sqrt{4}}\left(\frac{\theta_1(z,\tau)}{\eta^3(\tau)}\right)^2,\qquad  B_m=\sin[\frac{\pi m}{4}]\frac{\theta_1(z+\frac{m}{4},\tau)}{\theta_1(\frac{m}{4},\tau)}\frac{\theta_1(z,\tau)}{\eta^3},\quad m=1,2,3.
\end{eqnarray}
Explicit calculation via \eqref{eq:bjDef} yields $b_0=0$, $b_1=- \frac{i}{\sqrt{2}} $, $b_2= -i$, and $b_3=- \frac{i}{\sqrt{2}}$. Hence
\begin{equation}
\boxed{\begin{split}
    \mathcal{H}_1[\mathrm{HS}_4]&=-\frac{i}{2} \Big(\frac{\theta_1(z+\frac{1}{4},\tau)}{\theta_1(\frac{1}{4},\tau)}\frac{\theta_1(z,\tau)}{\eta^3}+2\frac{\theta_1(z+\frac{2}{4},\tau)}{\theta_1(\frac{2}{4},\tau)}\frac{\theta_1(z,\tau)}{\eta^3}+\frac{\theta_1(z+\frac{3}{4},\tau)}{\theta_1(\frac{3}{4},\tau)}\frac{\theta_1(z,\tau)}{\eta^3}\Big)\\
    &=y^{-1}-y+\mathcal{O}(q).
    \end{split}}
\end{equation}

For $\mathrm{HS}_6$ we have
\begin{eqnarray}
    Z[\mathrm{HS}_6]=\sum_{j=0}^5 B_j \overline{B_j},
\end{eqnarray}
with 
\begin{equation}
  B_0= \frac{1}{\sqrt{6}}\left(\frac{\theta_1(z,\tau)}{\eta^3(\tau)}\right)^2,\qquad  B_m=\sqrt{\frac{2}{3}}\sin[\frac{\pi m}{6}]\frac{\theta_1(z+\frac{m}{6},\tau)}{\theta_1(\frac{m}{6},\tau)}\frac{\theta_1(z,\tau)}{\eta^3},\quad m=1,2,\dots,5.
\end{equation}
Explicit calculation via \eqref{eq:bjDef} gives $b_0=0$, $b_1=-i\sqrt{\frac{1}{6}}$, $b_2=-i\frac{1}{\sqrt{2}} $, $b_3=-i\sqrt{\frac{2}{3}}$, $b_4=-i\frac{1}{\sqrt{2}}$, and $b_5= -i\sqrt{\frac{1}{6}}$. Hence
\begin{equation}
\boxed{\begin{split}
\mathcal{H}_1[\mathrm{HS}_6]&=-\frac{i}{6}  \Big (\frac{\theta_1(z+\frac{1}{6},\tau)}{\theta_1(\frac{1}{6},\tau)}\frac{\theta_1(z,\tau)}{\eta^3}+3\frac{\theta_1(z+\frac{2}{6},\tau)}{\theta_1(\frac{2}{6},\tau)}\frac{\theta_1(z,\tau)}{\eta^3}+4\frac{\theta_1(z+\frac{3}{6},\tau)}{\theta_1(\frac{3}{6},\tau)}\frac{\theta_1(z,\tau)}{\eta^3}\\
      &\qquad\quad +3\frac{\theta_1(z+\frac{4}{6},\tau)}{\theta_1(\frac{4}{6},\tau)}\frac{\theta_1(z,\tau)}{\eta^3}+\frac{\theta_1(z+\frac{5}{6},\tau)}{\theta_1(\frac{5}{6},\tau)}\frac{\theta_1(z,\tau)}{\eta^3}\Big)=y^{-1}-y+\mathcal{O}(q).
\end{split}}
\end{equation}

\section{Application to black hole microstate counting}
\label{sec:asy degeneracy}

\subsection{Asymptotic degeneracies on the boundary}
We would like to compute the asymptotics of $\hat{c}_1(\Delta,\ell)$ for $\Delta,\ell\to\infty$. Since we know its generating function, our problem is a standard exercise in analytic combinatorics.

\subsubsection{Saddle-point analysis}\label{subsubsec:saddle}

Extracting the coefficients of $\mathcal{H}_1(q,y)$ via contour integration we have
\begin{equation}
\label{integrand}
\tilde{d}(n,N,j)=\hat{c}_1(Nn,j)\simeq \int_0^1 \mathrm{d}\tau \int_0^1
\mathrm{d}z\ e^{-2\pi i  Nn\tau-2\pi i j z}\
\mathcal{H}_1(q,y).
\end{equation}
 
As in \cite{Sen:2012cj} we anticipate that at the saddle point value of $\tau$
is parametrically small\footnote{We have checked via the estimates in Appendix~\ref{app:thetas} that considering $\tau$ with real part parametrically close to rationals does not improve our results.} but that of $z$ is not. Taking $\mathrm{Re}z$ fixed inside $(0,1)$, let us parametrize the small-$\tau$ asymptotic of $\mathcal{H}_1$ with three coefficients $a_{0,1,2}$ as
\begin{equation}
\label{small tau asymptotic}
    \mathcal{H}_1(q,y)\approx e^{-\frac{2\pi i\, a_2 \, z^2}{\tau}+\frac{2\pi i\, a_1 \, z}{\tau}-\frac{2\pi i\, a_0}{\tau}}\, \tau,
\end{equation}
where we have multiplied the exponential by $\tau$ since this is the behavior that will arise in the $\mathrm{HS}_{2,3,4,6}$ cases below. Then
\begin{equation}
\tilde{d}(n,N,j)\sim \int_0^1 \mathrm{d}\tau \int_0^1
\mathrm{d}z\ e^{-2\pi i  Nn\tau-2\pi i j z-2\pi i
z^2\, a_2/\tau+2\pi i z\, a_1/\tau-2\pi i\, a_0/\tau} \tau.\label{eq:last}
\end{equation}

Extremizing the integrand we find the saddle point lies at
\begin{equation}
z_0=\frac{a_1}{2a_2}-\frac{j}{2a_2}\tau_0,\ \ \quad
\tau_0=\frac{i\sqrt{a_1^2-4a_0 a_2}}{2\sqrt{a_2\, Nn-\frac{j^2}{4}}}.
\end{equation}
The maximized exponent then turns out to have real part
\begin{equation}
S_\text{index}=2\pi\sqrt{Nn\, a_2-\frac{j^2}{4}}\cdot\frac{\sqrt{a_1^2-4a_0 a_2}}{a_2}.\label{eq:entropyB012}
\end{equation}
The saddle-point entropy also has an imaginary piece $-i\pi j a_1/a_2$, which implies that $\tilde d$ has phase oscillations even on the saddle point. In light of recent progress on complex saddles of higher-dimensional indices (see \emph{e.g.}~\cite{Choi:2018hmj,Choi:2018vbz,Cabo-Bizet:2018ehj,Benini:2018ywd}), we interpret this imaginary piece as a signal that bose-fermi cancellations have not been fully obstructed on the saddle-point. We will comment more on this in the discussion section.

Note that since the central charge of the supersymmetric sigma models $\mathrm{sym}^N \mathrm{HS}$ of our interest is $c=6N,$ the Cardy (and hence the bulk
Bekenstein-Hawking) entropy is $S_{\mathrm{BH}}=S_{\mathrm{Cardy}}=2\pi\sqrt{ \frac{cn}{6}-\frac{j^2}{4}}=2\pi\sqrt{ Nn-\frac{j^2}{4}}$. So it follows from \eqref{eq:entropyB012} that
\begin{equation}
    \text{in cases where }a_2=1,\qquad S_\text{index}=\sqrt{a_1^2-4a_0 }\,S_{\mathrm{BH}}.
\end{equation}

\subsubsection*{Examples}

The values of $a_{1,2}$ are given in Table~\ref{tab:B1B2} for the four HS cases of our interest, in the range where $a_0=0$ as we now explain.

For simplicity of exposition, we take $z\in\mathbb{R}$ and $\tau\in i\,\mathbb{R}_{>0}$, and  assume analytic continuation to more general $z,\tau$ gives the correct result. We have checked that a careful analysis justifies this assumption.

In the $\mathrm{HS}_2$ case, using the estimates in Appendix~\ref{app:thetas} we find the small-$\tau$ asymptotic of \eqref{eq:H1Z2result} to be (note that periodicity in $z$ implies we can focus on $z\in [0,1)$)
\begin{equation}
    \mathcal{H}_1[\mathrm{HS}_2]\approx \tau\, e^{\frac{i\pi}{\tau}[\vartheta(z+1/2)-\vartheta(1/2)+\vartheta(z)]} =\begin{cases}
    \tau\, \exp[\frac{i\pi}{\tau}(-2z^2+z)],\quad &z\in(0,1/2),\\
    \tau\,\exp[\frac{i\pi}{\tau}(-2z^2+3z-1)],\quad &z\in(1/2,1).
    \end{cases}
\end{equation}
Hence for the range $z\in(0,1/2)$ we have $a_0=0$ and $a_{2}=2a_1=1$ as in Table~\ref{tab:B1B2}, while for the range $z\in(1/2,1)$ we have instead $a_0=1/2,\, a_1=3/2,\, a_2=1.$ Importantly though, the entropy following from Eq.~\eqref{eq:entropyB012} would be the same whether we compute it with the $a_{0,1,2}$ that we get for $z\in(0,1/2)$ or the ones for $z\in(1/2,1)$. Therefore we have not bothered to record the differing coefficients in the range $z\in(1/2,1)$ in Table~\ref{tab:B1B2}.

In the $\mathrm{HS}_{3}$ case, using the estimates in Appendix~\ref{app:thetas} we find the small-$\tau$ asymptotic of \eqref{eq:H1Z3result} to be
\begin{equation}
\begin{split}
\mathcal{H}_1[\mathrm{HS}_3]&\approx\tau\, e^{\frac{i\pi}{\tau}[\vartheta(z+1/3)-\vartheta(1/3)+\vartheta(z)]}+ \tau\, e^{\frac{i\pi}{\tau}[\vartheta(z+2/3)-\vartheta(2/3)+\vartheta(z)]}\\
&\approx \tau\, e^{\frac{i\pi}{\tau}\cdot\mathrm{Max}[\vartheta(z+1/3)-\vartheta(1/3)+\vartheta(z),\vartheta(z+2/3)-\vartheta(2/3)+\vartheta(z)]} \\
&=\begin{cases}
    \tau\,\exp[\frac{i\pi}{\tau}(-2z^2+\frac{4}{3}z)],\quad &z\in(0,1/2) ,\\
    \tau\,\exp[\frac{i\pi}{\tau}(-2z^2+\frac{8}{3}z-\frac{2}{3})] ,\quad &z\in(1/2,1).
    \end{cases}
\end{split}
\end{equation}
Hence for the range $z\in(0,1/2)$ we have $a_0=0$ and $a_{2}=\frac{3}{2}a_1=1$ as in Table~\ref{tab:B1B2}, while for the range $z\in(1/2,1)$ we have instead $a_0=1/3,\, a_1=4/3,\, a_2=1.$ Importantly though, the entropy following from Eq.~\eqref{eq:entropyB012} would be the same whether we compute it with the $a_{0,1,2}$ that we get for $z\in(0,1/2)$, or the ones for $z\in(1/2,1)$.

Similarly, in the $\mathrm{HS}_{4,6}$ cases the coefficients $a_{0,1,2}$ differ for $z\in(0,1/2)$ and $z\in(1/2,1)$. However, the resulting entropies are the same. Hence we only consider the range $z\in(0,1/2)$ where $a_0=0$.\\

\begin{table}[h]
    \centering
    \begin{tabular}{|l|c|c|c|}
    \hline
       HS  & $a_2$ & $a_1$ & $S$\\
       \hline
       HS$_2$  & $1$ & $\frac{1}{2}$ & $S_{\mathrm{BH}}/2$\\
        HS$_3$ & $1$&$\frac{2}{3}$ & $2S_{\mathrm{BH}}/3$\\
        HS$_4$ & $1$&$\frac{3}{4}$ & $3S_{\mathrm{BH}}/4$\\
        HS$_6$ &$1$&$\frac{5}{6}$ & $5S_{\mathrm{BH}}/6$\\
        \hline
    \end{tabular}
    \caption{The coefficients $a_1,a_2$ parameterize the small $\tau$ asymptotics of $\mathcal{H}_1$ as in \eqref{small tau asymptotic}, for $z\in(0,1/2)$ where $a_0=0$. The resulting entropies are compared with the Bekenstein-Hawking entropy $S_{\mathrm{BH}}=2\pi\sqrt{ Nn-\frac{j^2}{4}}$.}
    \label{tab:B1B2}
\end{table}

\subsubsection*{Logarithmic corrections}

To extract the logarithmic term, it is convenient to introduce a scale parameter $\Lambda\to\infty$ \cite{Sen:2012cj} and scale the charges as
\begin{equation}
    n \sim \Lambda, \quad c=6N\sim \Lambda^2,\quad j \sim \Lambda^{\frac{3}{2}}.
\end{equation}
The saddle point for all the hyperelliptic cases is written compactly as
\begin{equation}
z_0=\frac{a_1}{2a_2}-\frac{J}{2a_2}\tau_0 \sim \Lambda^0,\ \ \quad
\tau_0=\frac{i\sqrt{a_1^2-4a_0 a_2}}{2\sqrt{a_2\, Nn-\frac{j^2}{4}}} \sim \Lambda^{\frac{-3}{2}}.\label{eq:z&tauSaddleScaling}
\end{equation}

To find the logarithmic correction, we also need to find all the $\Lambda$ dependence in the integrand of 
\begin{equation}
\tilde{d}(n,N,j)\sim \int_0^1 \mathrm{d}\tau \int_0^1
\mathrm{d}z\ e^{-2\pi i  Nn\tau-2\pi i j z-2\pi i
z^2\, a_2/\tau+2\pi i z\, a_1/\tau-2\pi i\, a_0/\tau} \tau.
\end{equation} 
We know from \eqref{eq:z&tauSaddleScaling} that $\tau_0\sim \Lambda^{-\frac{3}{2}}$. By taking the second derivatives of the exponent of the integrand above we find the standard deviations around the saddle point to scale as
\begin{equation}
    \Delta z\propto\frac{1}{\sqrt{1/\tau_0}}\sim
    \Lambda^{-\frac{3}{4}}, \quad  \Delta \tau\propto\frac{1}{\sqrt{1/\tau_0^3}}\sim\Lambda^{-\frac{9}{4}}\,.
\end{equation}
Here $\Delta \tau$ and $\Delta z$ represent the effective width of integration. Then the power of $\Lambda$ arising from the one-loop correction to the saddle-point result is
\begin{equation}
    \sim\tau_0\,\Delta z\,\Delta\tau\sim\Lambda^{\frac{-9}{2}}= \mathrm{exp}\big[{-9} \log\Lambda^{1/2}\big].
\end{equation}
The logarithmic correction to entropy is hence 
\begin{equation}
-9\log\Lambda^{1/2}.\label{eq:saddleLogMicro}
\end{equation}

\subsubsection{Rademacher expansion for $k=2$}\label{subsec:Rademacher}

Let us now focus on the $\mathrm{HS}_2$ case. Recall the counting function $\mathcal{H}_1[\mathrm{HS}_2](q,y)$ is given by
\begin{equation}
\mathcal{H}_1(z,\tau)=-2i\,\frac{\theta_2(z,\tau)}{\theta_2(\tau)}\frac{\theta_1(z,\tau)}{\eta^3(\tau)}.
\end{equation}
It turns out $\mathcal{H}_1(z,\tau/2)$ is a weak Jacobi form of weight $-1$ and index $2$:
\begin{equation}
    \mathcal{H}_1(z,\tau/2)=-i\,\phi_{-1,2}(\tau,z).\label{eq:HS2wjf}
\end{equation}
We discovered this by comparing the Fourier coefficients and checking
\begin{equation}
\label{identity}
-2i\,\frac{\theta_2(z,\tau/2)}{\theta_2(\tau/2)}\frac{\theta_1(z,\tau/2)}{\eta^3(\tau/2)}=(- i)\frac{\theta_1(2z,\tau)}{\eta^3(\tau)}=(-i)\phi_{-1,2}(\tau,z).
\end{equation}
This identity can indeed be proven using (see Appendix~A.3 in \cite{Takayanagi:2022xpv})
\begin{equation}
    \frac{\theta_1(z,2 \tau)}{\eta(2 \tau)}= \prod_{k=0}^{1} \frac{\theta_1(\frac{z}{2}+\frac{k}{2}, \tau)}{\eta(\tau)},
\end{equation}
together with $2 \eta^2(2 \tau)=\theta_2(\tau) \eta(\tau).$

Details on weak Jacobi forms can be found in \cite{Dabholkar:2012nd,Ferrari:2017msn}. 
 A weak Jacobi form has a double Fourier expansion as
 \begin{eqnarray}
     \phi^{}_{k,m}(\tau,z)=\sum_{\tilde n,r\in \mathbb{Z}} c(\tilde n,r) q^{\tilde n} y^r,
 \end{eqnarray}
such that $c(\tilde n,r)=0$ unless $\tilde n\geq 0$. This can be verified for $\phi_{-1,2}$ given above by doing the explicit expansion in $q$ and $y$ with Mathematica. Weak Jacobi forms moreover have the periodicity property
\begin{equation}
     c(\tilde n,r)=C_{\nu}(4 \tilde n m -r^2),
\end{equation}
where $C_{\nu}(\delta)$ depends  on  $\nu:= r\,\,  \mathrm{mod}\,\, 2m,$ as well as on $\delta:=4 \tilde n m -r^2$.
 
We have a Jacobi form of weight $k=-1$ and index $m=2$. So $\delta=8 \tilde n-r^2$ and $\nu= r \, \mathrm{mod} \, 4$.  Hence we have to consider four functions $C_{\nu},$ with  $\nu=0,1,2,3$. For our specific weak Jacobi form it turns out $\phi_{-1,2}(\tau,z)=-\phi_{-1,2}(\tau,-z),$ which implies
\begin{equation}
    \begin{split}
        C_3(\delta)&=-C_1(\delta),\\
        C_0(\delta)&=C_2(\delta)=0.\label{eq:HS2period}
    \end{split}
\end{equation}
We hence focus on $\nu=1$ and denote $C_1(\delta)$ by $C(\delta)$ for simplicity. The Hardy-Ramanujan-Rademacher expansion gives (see \emph{e.g.}~\cite{Ferrari:2017msn,Murthy:2015zzy})
\begin{equation}
    C(\delta)= 2 \pi (\frac{\pi}{2})^{\frac{5}{2}}\sum_{\mathrm{c}=1}^{\infty} \mathrm{c}^{\frac{-7}{2}} K_\mathrm{c}(\delta) \tilde {I}_{5/2}(\frac{\pi \sqrt{\delta}}{2 \mathrm{c}}).
\end{equation}
Here $\tilde {I}_{5/2}$ is a modification of the Bessel-I function
\begin{equation}
  \tilde{I}_{\rho}(z)=(\frac{z}{2} )^{-\rho} I_{\rho}(z)=\frac{1}{2 \pi i} \int _{\epsilon-i \infty}^{\epsilon+i \infty}\frac{d \sigma}{\sigma^{\rho+1}} \, \mathrm{exp}(\sigma+\frac{z^2}{4 \sigma}) .
\end{equation}
For $\mathrm{c}=1$ the Kloosterman sum $K_\mathrm{c}=1$, while for $\mathrm{c}>1$ there is a complicated expression (see \emph{e.g.}~\cite{Iliesiu:2022kny}). Now the degeneracy is simply
\begin{equation}
    \frac{1}{2 \pi} (\frac{2}{\pi})^{\frac{5}{2}} C(\delta)=\sum_{\mathrm{c}=1}^{\infty}\mathrm{c}^{\frac{-7}{2}} K_\mathrm{c}(\delta) \tilde {I}_{5/2}(\frac{\pi \sqrt{\delta}}{2\mathrm{c}}).
\end{equation}
To find the asymptotic at large $z \rightarrow \infty$, we just need to expand the Bessel I function. The leading and sub-leading order contributions are found from the $\mathrm{c}=1$ term as
\begin{equation}
\label{eq:entropyRademacher}
    \log[C(\delta)]=\frac{\pi}{2}  \sqrt{\delta}- \frac{3}{2}\mathrm{\log}[\delta]+\mathrm{log[\frac{ \pi^{\frac{7}{2}}}{2^{\frac{3}{2}}}]}+O(\frac{1}{\sqrt{\delta}}).
\end{equation}

Since we are actually interested in $\mathcal{H}_1(z,\tau)=-i\phi_{-1,2}(2\tau,z)$, and since the power of $q$ in the expansion of the latter is $2\tilde n$, we set $n=2\tilde n.$ We are thus interested in $C(4 Nn-r^2)$, the leading entropy of which follows from \eqref{eq:entropyRademacher} to be
\begin{equation}
    S_{\text{index}}= \frac{\pi}{2}\sqrt{4  Nn-j^2}= \pi \sqrt{  Nn-\frac{j^2}{4}}.
\end{equation}

To extract the subleading logarithmic contribution we follow \cite{Sen:2012cj} and consider the scaling limit $N \sim \Lambda^2,\ n \sim \Lambda^1,\  j\sim \Lambda^{\frac{3}{2}}$. After the scaling, the $\log\Lambda$ term in \eqref{eq:entropyRademacher} becomes $-\frac{9}{2}\log\Lambda $.

Note that both the leading and the subleading logarithmic result are compatible with the ones found above via the saddle-point analysis. They are also compatible with numerics, as we discuss next.

\subsubsection{Numerical check}

The Fourier coefficients of $\mathcal{H}_1[\mathrm{HS}]$ can be numerically evaluated. They are defined via
\begin{equation}
\mathcal{H}_1[\mathrm{HS}]=\sum_{n,l\in \mathbb{Z}} c(n,l) q^n y^l.
\end{equation}
Expanding the explicit $\mathcal{H}_1[\mathrm{HS}_k]$ functions above, via Mathematica for instance, one finds that $c(n,l)$ is actually only a function of $\delta:=4n-l^2$ and $\nu:=l\ \text{mod}\ 2k$ (we hope the $k$ which we use in this subsection to label various $\mathrm{HS}_k$ does not get confused with the weight of weak Jacobi forms denoted $k$ in the previous subsection):
\begin{equation}
    c(n,l)=C^{}_{l\, \text{mod}\, 2k}(4  n-l^2).
\end{equation}
This periodicity, combined with the fact that $\mathcal{H}_1[\mathrm{HS}_k](z,\tau)=-\mathcal{H}_1[\mathrm{HS}_k](-z,\tau)$, implies
\begin{equation}
    \begin{split}
        C_{-\nu}(\delta)&=-C_{\nu}(\delta),\\
        C_0(\delta)&=C_k(\delta)=0.
    \end{split}
\end{equation}
Moreover, only $C_\nu(-1)$ is nonzero among $C_\nu(\delta<0)$. This suggests that $\mathcal{H}_1[\mathrm{HS}_k](z,\tau/k)$ might be a component of a $k$-dimensional vector-valued weak Jacobi form of index $k$, and we indeed found this form explicitly in the $\mathrm{HS}_2$ case.\footnote{For $\mathrm{HS}_2$ because odd powers of $q$ are absent in $\mathcal{H}_1(z,\tau)$, the function $\mathcal{H}_1(z,\tau/2)$ ends up being single-valued, so instead of a $k=2$ dimensional vector-valued weak Jacobi form we get a single weak Jacobi form. For $k=3,4,6$ we have checked that similar reductions do not occur, so actual $k$-dimensional vectors arise.}
 
\begin{figure}[!tbp]
  \centering
  \subfloat[$\mathrm{HS}_2$ ]{\includegraphics[scale=0.45]{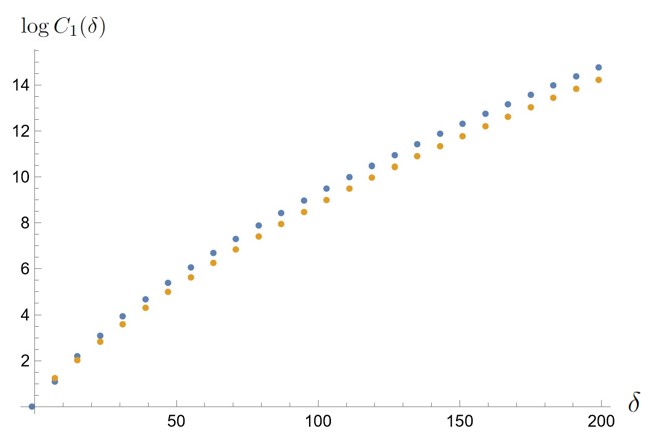}
	\label{f:hs2}}
  \hfill
  \subfloat[$\mathrm{HS}_3$]{\includegraphics[scale=0.45]{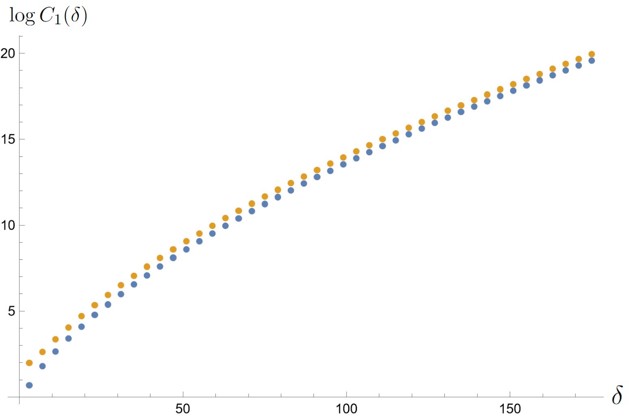}
	\label{f:hs3}}\\
  \subfloat[$\mathrm{HS}_4$ ]{\includegraphics[scale=0.45]{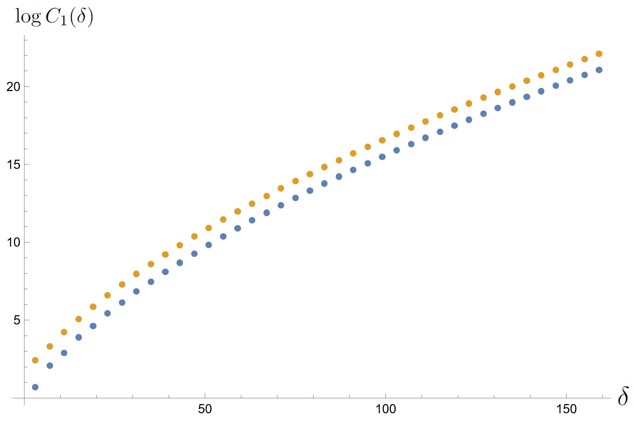}
	\label{f:hs4}}
  \hfill
  \subfloat[$\mathrm{HS}_6$]{\includegraphics[scale=0.45]{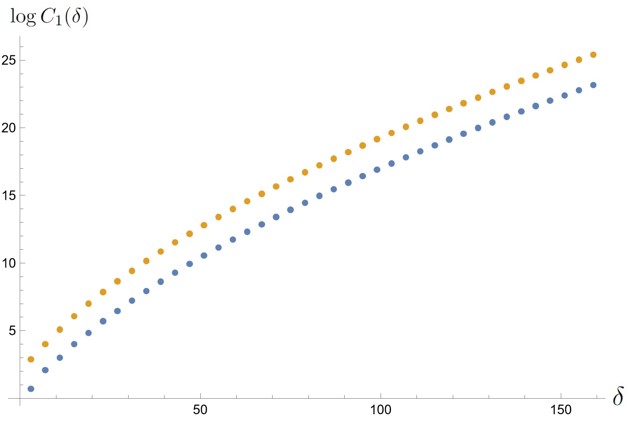}
	\label{f:hs6}}
  \caption{The orange data points are saddle-point approximations, while the blue data points are the numerical values found via Mathematica.}\label{Fig:numerics}
\end{figure}

We have plotted $\log C_1(\delta)$  as found via Mathematica, along with the saddle-point expectation in Figure~\ref{Fig:numerics}. The saddle-point expectation for log of the degeneracy  is $a_1\pi \sqrt\delta-\frac{3}{2} \log\delta$. For various cases, the coefficient $a_1$ is listed in Table~\ref{tab:B1B2}. To obtain the numerical plots, we have fixed the value $l=1$, read the coefficients of $q^n\,y^1,$ plotted the log of their absolute value against $4n-(1)^2$ for different values of $n.$ We have also checked that considering other values of $l$ with nonzero $c(n,l)$ leads to similar plots.

\subsection{Comparison with bulk entropies}\label{subsec:bulkEntropies}

Our AdS$_3$/CFT$_2$ dual pairs arise from low-energy limits of back-reacting D1-D5 systems. We spell out some of the details below, closely following Sections 3.2 and 4.1 of \cite{ArabiArdehali:2021iwe}.

\subsection*{The D1-D5 system}

 The non-backreacted geometry is $\mathbb{R}\times S^1\times \mathbb{R}^2\times (\mathbb{C}\times T^4)/G$. One can  wrap $Q_1$ D1-branes on the $\mathbb{R}\times S^1$ part of the geometry, and wrap $Q_5$ D5-branes on $\mathbb{R}\times S^1\times \mathcal{D}_4$, where $\mathcal{D}_4\sim T^4/G$ is a four-real-dimensional divisor inside $(\mathbb{C}\times T^4)/G.$

After including the backreaction, we have a p-brane solution in IIB supergravity with the following metric, dilaton, and
3-form flux (see \emph{e.g.} \cite{Dabholkar:1997rk}, and note that the difference in $F_3$ between the factors $e^{-2\phi}$ there and $e^{+2\phi}$ here is due to the difference between $\ast_6$ and $\ast_{10}$):
\begin{equation}
\begin{split}
e^{-2\phi}&=f_5/f_1,\\
\mathrm{d}s^2&=f_1^{-1/2}f_5^{-1/2}\mathrm{d}x_{||}^2+f_1^{1/2}f_5^{1/2}(\mathrm{d}r^2+r^2\mathrm{d}\Omega_3^2)+f_1^{1/2}f_5^{-1/2}\mathrm{d}x_{T^4}^2,\\
F_3&=2r_5^2\epsilon_3+2r_1^2 e^{+2\phi}\ast_{10}\epsilon_7,\\
f_i&:=1+r_i^2/r^2\quad\ i=1,5,\label{eq:pBrane}
\end{split}
\end{equation}
where $\mathrm{d}x_{||}^2=-\mathrm{d}t^2+\mathrm{d}x^2$, and $x$ is
the coordinate along the D1-branes. The radial
coordinate of $\mathbb{R}^2\times \mathbb{C}$ is $r$. The volume forms $\epsilon_3$ and
$\epsilon_7$ are the volume forms of a three-cycle $\mathcal{C}_3$ and the
seven-cycle $\mathcal{C}_7$ respectively, at $r=1$ inside $\mathbb{R}^2\times( \mathbb{C}\times T^4)/G$.

One can integrate $F_3$ at $r\to\infty$ in order to relate the local parameters $r_i$ to the global charges $Q_i$ of
the D-branes. We find
\begin{equation}
Q_5=\frac{1}{4\pi^2\alpha'
g}\int_{\mathcal{C}_3}F_3=\frac{r_5^2}{\alpha' g},\quad
Q_1=\frac{1}{(4\pi^2\alpha')^3
g}\int_{\mathcal{C}_7}\ast_{10}F_3=\frac{r_1^2 v}{\alpha'
g},\label{eq:fluxIntegration}
\end{equation}
with $g$ the closed string coupling. To evaluate the integrals in
(\ref{eq:fluxIntegration}) we have used
\begin{equation}
\int
\epsilon_7=\frac{\mathrm{vol}(S^3)\times\mathrm{vol}(T^4)}{|G|},\quad
\mathrm{\ and}\quad\ \int \epsilon_3=\mathrm{vol}(S^3).\label{eq:volFormulas}
\end{equation}
Therefore, we have
\begin{equation}
r_5^2=g\alpha' Q_5,\quad\ r_1^2=\frac{g\alpha'
Q_1}{v},\label{eq:r1andr5}
\end{equation}
which is exactly like the standard D1-D5 systems with $G=1$, but now we have
\begin{equation}
v:=\frac{\mathrm{vol}(T^4)/((2\pi)^4 \alpha'^2)}{|G|},\label{eq:v}
\end{equation}
where $|G|=k$ is the order of the group  $G=\mathbb{Z}_k$. 
See Section~3.2 of \cite{ArabiArdehali:2021iwe} for more details.\\

 One can also calculate the value of the
3d Newton's constant in the AdS$_3$ space which arises from the
decoupling limit of the p-brane solution (\ref{eq:pBrane}) . The
near-horizon limit of the geometry is \cite{Maldacena:1997re} (see also \cite{Kraus:2006wn})
\begin{equation}
\mathrm{d}s_{10}^2=\alpha'(\mathrm{d}s_{\mathrm{AdS}_3}^2+\ell^2
\mathrm{d}\Omega_3^2)+\mathrm{d}x_{T^4_{\text{n.h.}}}^2,\label{eq:10dNHD1D5}
\end{equation}
where
\begin{equation}
\mathrm{d}s_{\mathrm{AdS}_3}^2=\frac{U^2}{\ell^2}\mathrm{d}x_{||}^2+\frac{\ell^2}{U^2}\mathrm{d}U^2,\qquad
U=\frac{r}{\alpha'},\qquad \ell=\left(\frac{g^2 Q_1
Q_5}{v}\right)^{1/4},\qquad \mathrm{d}x_{T^4_{\text{n.h.}}}^2=\sqrt{\frac{ Q_1}{v Q_5}}\mathrm{d}x_{T^4}^2. \label{eq:nhGeometry}
\end{equation}
Note that according to \eqref{eq:10dNHD1D5} the near-horizon $S^3$ has radius $R_{S^3}=\sqrt{\alpha'}\ell$, and should not be confused with the unit $S^3$ used in Eq.~\eqref{eq:volFormulas}.

The solution \eqref{eq:10dNHD1D5} is true for any value of $v$, including ours which is smaller than the standard $\mathcal{N}=(4,4)$ value by a factor of $|G|$. The reason is that the 10d supergravity equations are local with field strengths $F_3$ and $*_{10} F_3$. 
We can then compute in the near-horizon limit
\begin{equation}
G^N_{\mathrm{AdS_3}}=\frac{G^N_{10}\cdot e^{2 \phi}}{\mathrm{vol}((S^3 \times
T^4)/G)}\bigg|_{\text{n.h.}}=\frac{8\pi^6 g^2 \alpha'^4 \cdot r_1^2/r_5^2}{2\pi^2 R_{S^3}^3\times (2\pi)^4
\alpha'^2 v \frac{ Q_1}{v Q_5} }=\frac{\sqrt{\alpha'}(g^2 Q_1 Q_5/v)^{1/4}}{4Q_1
Q_5}.\label{eq:GAdS3}
\end{equation}
The extra factor of $e^{2 \phi}$ in the numerator came from the transformation of the string frame to the Einstein frame (see for example Eq.~(4.8) of \cite{Kraus:2006wn}).

From (\ref{eq:nhGeometry}) and
(\ref{eq:GAdS3}) the Brown-Henneaux central charge \cite{Brown:1986nw} becomes
\begin{equation}
c_{0}=3R_{\mathrm{AdS}_3}/2G^N_{\mathrm{AdS_3}}=6Q_1
Q_5,\label{eq:BHcc}
\end{equation}
where we have used
$R_{\mathrm{AdS}_3}(=R_{S^3})=\sqrt{\alpha'}\ell$. The upshot of this calculation is that the expression for $c_0$ in terms of $Q_1$ and $Q_5$ is exactly the same as that of the standard
D1-D5 systems with $G=1$. (A \emph{boundary} derivation of $c_0$ confirming the bulk value \eqref{eq:BHcc} was given in \cite{ArabiArdehali:2021iwe} based on the relation $c_0=\frac{3}{2}\mathrm{dim}\mathcal{M}^{Q_5}_{Q_1}$ between the sigma model central charge $c_0$ and the dimension $\mathrm{dim}\mathcal{M}^{Q_5}_{Q_1}=4Q_1Q_5$ of the moduli space of D1 instantons inside the D5 branes.)

In summary, on the one hand $G_{\mathrm{AdS}_3}^N$ increases by a factor of $|G|$ compared to the $G=1$ case because the volume of   $(S^3 \times T^4)/G$ used in \eqref{eq:GAdS3} is reduced by $|G|$; on the other hand, to have the same $Q_1$ flux, we need greater field strength $\ast_{10}F_3$ as in \eqref{eq:fluxIntegration}, which implies larger $r_1$ and therefore larger $R_{\mathrm{AdS}_3}=\sqrt{\alpha'}\ell$ in the near-horizon limit. Then the ratio, the central charge $c_0$, for the same $Q_1,Q_5$ remains unaffected by non-trivial $G$.

Since AdS/CFT identifies the Brown-Henneaux central charge $6Q_1Q_5$ with the CFT $\mathrm{sym}^N\mathrm{HS}$ central charge $6N$, we have
\begin{equation}
    Q_1Q_5=N.\label{eq:ccDic}
\end{equation}

\subsection*{The D1-D5-P system and its Bekenstein-Hawking entropy}

One can also add $n$ units of left-moving momentum along the $S^1$. This excited system is called the D1-D5-P system, and has the metric (see \emph{e.g.} \cite{Dabholkar:1997rk})
\begin{equation}
\begin{split}
f_1^{-1/2}f_5^{-1/2}\big(-\mathrm{d}t^2+\mathrm{d}x^2+(f_n-1)(\mathrm{d}t-\mathrm{d}x)^2\big)+f_1^{1/2}f_5^{1/2}(\mathrm{d}r^2+r^2\mathrm{d}\Omega_3^2)+\mathrm{d}x_{T^4_r}^2,\label{eq:pBraneExcited}
\end{split}
\end{equation}
with 
\begin{equation}
f_i=1+r_i^2/r^2,\ i=1,5,\qquad f_n:=1+r_n^2/r^2,\quad r_n^2=\frac{g^2\alpha'^2
n}{vR_{S^1}^2},\qquad \mathrm{d}x_{T^4_r}^2=f_1^{1/2}f_5^{-1/2}\mathrm{d}x_{T^4}^2\label{eq:rn}
\end{equation}
where $R_{S^1}$ is the radius of the circle.

The effect of $G$ is twofold: $i)$ it changes the ranges of various coordinates due to identifications on $S^3\times T^4$ by $G$, and $ii)$ the relation between $r^2_1$ and $Q_1$, and the one between $r^2_n$ and $n$, change because they contain $v$ which depends on $|G|$ as in (\ref{eq:v}).

While one can reduce the Strominger-Vafa black brane on $S^1\times T^4$ to yield a lower-dimensional picture as a 5d black hole, in our case $G$ mixes the $T^4$ with the $S^3$, so the only smooth lower-dimensional picture is obtained by reducing only on the $S^1$, which gives a 9d black brane.

The entropy of the 9d Strominger-Vafa black brane is given by the Bekenstein-Hawking formula (\emph{cf}. \cite{Strominger:1996sh,David:2002wn})
\begin{equation}
\begin{split}
    S_{\mathrm{SV}}=\frac{A}{4G_9\cdot e^{2\phi}}\bigg|_{r=r_n}=\frac{\mathrm{vol}(S^3)\,\mathrm{vol}(T^4_r)}{4G_9\cdot e^{2\phi}}\bigg|_{r=r_n}=\frac{(2\pi^2 R_h^3)\mathrm{vol}(T^4)}{4G_9}&=\frac{(2\pi^2 r_1r_5r_n)\mathrm{vol}(T^4)}{4G_9}\\
&=2\pi\sqrt{Q_1Q_5n},\label{eq:Ssv}
\end{split}
\end{equation}
where we have used the fact that the horizon is at $r=r_n$, and in going to the second line we have used $G_9=G_{10}/2\pi R_{S^1}=8\pi^6 g^2\alpha'^4/2\pi R_{S^1}$ for the 9d Newton constant.\\

For our 9d black branes, the expression for the entropy in terms of the metric parameters $r_{1,5,n}$ is smaller by a factor of $|G|$, because the orbifolding reduces the horizon area. However, according to (\ref{eq:r1andr5}) and (\ref{eq:rn}), when we write $r_1$ and $r_n$ in terms of the global charges $Q_1$ and $n$, we have the factors of $\sqrt{|G|}$ which arise from the denominator of $v$ in (\ref{eq:v}). Hence analogously to (\ref{eq:Ssv}) in our case we have
\begin{equation}
\begin{split}
    S_{\mathrm{SV}/G}&=\frac{\left((2\pi^2 r_1r_5r_n)\mathrm{vol}(T^4)\right)/|G|}{4G_9}\\
    &=\frac{2\pi\sqrt{(|G|Q_1)Q_5(|G|n)}}{|G|}=2\pi\sqrt{Q_1Q_5n}.\label{eq:S2c2}
\end{split}
\end{equation}
So the BH-entropy in terms of the charges $Q_1,Q_5,n$ is exactly as it was in the standard (4,4) case of Strominger-Vafa where $G$ is trivial. This is not surprising. Using standard AdS$_3$/CFT$_2$ arguments we can argue that the entropy of the black branes is reproduced by the Cardy formula \cite{Strominger:1997eq}, which is actually fixed by the central charge. We found in (\ref{eq:BHcc}) that the leading central charge of the $(2,2)$ cases in terms of $Q_{1,5}$ are exactly the same as in the standard $(4,4)$ cases. So the formulas for the corresponding entropies in terms of $Q_{1,5}$ should be the same as well.

Addition of angular momentum shifts $Q_1Q_5n\to Q_1Q_5n-j^2/4$ \cite{Breckenridge:1996is} and hence modifies \eqref{eq:S2c2} to
\begin{equation}
    S_{\mathrm{BMPV}/G}=2\pi\sqrt{Q_1Q_5n-\frac{j^2}{4}},
\end{equation}
which using the AdS/CFT dictionary \eqref{eq:ccDic} becomes
\begin{equation}
    S_{\mathrm{BH}}=2\pi\sqrt{Nn-\frac{j^2}{4}},\label{eq:ent2c2rot}
\end{equation}
as we claimed in Section~\ref{subsubsec:saddle}.

To recap, while general AdS$_3$/CFT$_2$ arguments together with the Cardy formula imply that the Brown-Henneaux central charge (\ref{eq:BHcc}) captures the Bekenstein-Hawking entropy (\ref{eq:ent2c2rot}) in the $(2,2)$ cases of our interest as well, as displayed in Table~\ref{tab:B1B2} the protected boundary superconformal indices fail to reproduce the same result.

\subsection*{Logarithmic correction to the Bekenstein-Hawking entropy}

The one-loop correction to the effective action on the near-horizon region of the black brane system gives a logarithmic correction to the entropy.\footnote{While the sufficiency of near-horizon considerations is contested for asymptotically AdS black branes \cite{Liu:2017vbl}, it is widely accepted in the asymptotically flat situations such as ours.  See \emph{e.g.}~\cite{Sen:2011ba,Keeler:2014bra,Sen:2012dw,H:2023qko} and references therein.}

We consider the limit where the mass $M$ and charges $Q_i$ of the black brane are large while its temperature is kept zero:
\begin{equation}
    Q_i\sim \Lambda,\quad M \sim \Lambda,\qquad T=0. 
\end{equation}
Then the general formula for the logarithmic contribution to the entropy from a field with kinetic operator $\mathcal{A}$ can be written as (see \emph{e.g.} \cite{Sen:2012cj,Liu:2017vll})
\begin{equation}
    (-1)^F (\beta_\mathcal{A}-1)n^0_{\mathcal A}\log \Lambda^{1/2}. \label{eq:macLogMaster}
\end{equation}
The coefficients $\beta$ are given by the formulas
\begin{equation}
    \beta_{\mathrm{graviton}}=D/2,\quad \beta_{\mathrm{gravitino}}=D-1,\quad \beta_{A_1}=D/2-1,\label{eq:betas}
\end{equation}
where $D$ is the number of ``large'' dimensions, growing with $\Lambda$ in the near-horizon geometry. In the cases of our interest the near-horizon geometry is of the form $\mathrm{AdS}_2\times \frac{S^3\times T^4}{\mathbb{Z}_k}\times S^1$, with the $\mathrm{AdS}_2$ and $S^3$ parts growing as $\Lambda\to\infty$, so $D=5$. (For more details on the $k=1$ case see \cite{Sen:2012cj}.) The coefficients $n^0$ denote the effective number of AdS$_2$ zero-modes, and are given by (see \emph{e.g.}~\cite{Liu:2017vll})
\begin{equation}
    n^0_{\mathrm{graviton}}=-3,\quad n^0_{\mathrm{gravitino}}=-2,\quad n^0_{A_1}=-1,\label{eq:AdS2zeros}
\end{equation}
where $A_1$ indicates a one-form (alternatively gauge field, or vector).

Let us start by reviewing the $T^4$ calculation. The relevant 5d field content is summarized in Table~\ref{tab:T4spec}. We ignore scalars and fermions, and only keep differential forms, gravitinos, and the graviton, as they can potentially have zero modes. Using the formula~\eqref{eq:macLogMaster} we obtain the coefficient of $\log\Lambda^{1/2}$ in the logarithmic correction to be
\begin{equation}
\begin{split}
&\underbrace{(\frac{3}{2}-1)(-1\times 27)}_{\text{gauge fields}} +\underbrace{(\frac{5}{2}-1) (-3\times 1-1\times 4)}_{\text{metric}}
-\underbrace{(4-1)(-2\times 2)}_{\text{gravitino}}=-12.
\end{split}\label{eq:logcorrection5dT4}
\end{equation}
The 27 gauge fields arise from the 19 one-forms in 5d as seen in Table~\ref{tab:T4spec}, together with the 8 two-forms which we Hodge-dualize to one-forms. That the two-forms should be Hodge-dualized to one-forms has been related to an ensemble choice in \cite{Sen:2011ba}. For our purposes, the takeaway lesson is that to obtain the correct result \eqref{eq:logcorrection5dT4} compatible with microscopics \cite{Sen:2012cj}, one has to dualize two-forms to one-forms (since they are not equivalent quantum mechanically \cite{Duff:1980qv}), and that is what we will do also for the $\mathrm{HS}$ cases below. The metric piece contains besides the contribution from 1 metric, contributions from 4 gauge fields that arise from the one-form AdS$_2$ zero-modes that KK reduction of the 5d metric on a rotating $S^3$ (with 4-dimensional isometry $\mathrm{SU}(2)\times U(1)$) yields. Finally, the number of gravitinos follows from the amount of supersymmetry. For more details, see the derivation of Eq.~(4.16) in \cite{Sen:2012cj}.

\begin{table}[t]
\centering
\begin{tabular}{ |c|c|c|c|c| } 
 \hline
 10d & $g_{\mu\nu}$ & $B_{\mu\nu}$ & $C_{\mu\nu}$ & $C_{\alpha\beta\gamma\delta}^+$\\
 \hline\hline
9d & $g_{\mu\nu}$, $A_\mu$ & $B_{\mu}$, $B_{\mu\nu}$ & $C_{\mu},\ C_{
\mu\nu}$ &$C_{\alpha\beta\gamma}$ or $C_{\alpha\beta\gamma\delta}$ \\ 
 \hline\hline
5d & $g_{\mu\nu}$, $5A_\mu$ & $5B_{\mu}$, $B_{\mu\nu}$ & $5C_{\mu},\  C_{\mu\nu}$ &$4C_\alpha,\ 6C_{\alpha\beta}$ \\ 
 \hline
\end{tabular}
\caption{The first row contains  the bosonic fields of IIB SUGRA. The second row is the 9d field content after compactifying on the $S^1$; note that 10d self-duality of the four-form implies that in 9d we have \emph{either} the four-form or the three-form. The 5d field content after further compactification on $T^4$ is in the third row. The additional one-forms in going from 9d to 5d come from the 4 one-cycles of $T^4$, or in the case of $C_\alpha$ from $C_{\alpha\beta\gamma\delta}$ wrapping the 4 three-cycles of $T^4$. The 6 two-forms $C_{\alpha\beta}$ come from $C_{\alpha\beta\gamma\delta}$ wrapping the 6 two-cycles of $T^4$. The Hodge diamond of $T^4$ can be found in Appendix~\ref{app:hodge}. }\label{tab:T4spec}
\end{table}

Now on to the $\mathrm{HS}$ case. We adopt a low-energy perspective where only the macroscopic AdS$_2\times S^3/\mathbb{Z}_k$ part of the near-horizon geometry is visible. Then proceed as in the 5d calculation performed in \cite{ArabiArdehali:2021iwe} for the $\mathrm{ES}$ case.\footnote{In \cite{ArabiArdehali:2021iwe} 
for the $\mathrm{ES}$ case a 9d calculation of the log correction was presented as well. The correct Hodge duality frame in that case was inferred from comparison with the parent $K3$ case where a specific duality frame was found that led to the correct result. We do not pursue the 9d computation for the $\mathrm{HS}$ case here because we are not aware of a duality frame that yields the correct result \eqref{eq:logcorrection5dT4} via a 9d calculation (as in \cite{ArabiArdehali:2021iwe}) for the parent $T^4$ case.}

To obtain the macroscopic (5d) field content, we first reduce IIB SUGRA on the $S^1$, and then on $\mathrm{HS}$. The bosonic 5d field content is summarized in Table~\ref{tab:HS2spec}. Formula~\eqref{eq:macLogMaster} gives the coefficient of $\log\Lambda^{1/2}$ in the logarithmic correction as
\begin{equation}
\begin{split}
&\underbrace{(\frac{3}{2}-1)(-1\times 15)}_{\text{gauge fields}} +\underbrace{(\frac{5}{2}-1)\times (-3\times 1-1\times 2)}_{\text{metric}}
-\underbrace{(4-1)(-2\times 1)}_{\text{gravitino}}=-9.\\
\end{split}\label{eq:logcorrection5dHS}
\end{equation}
Note that again the 5d two-forms have been dualized to one-forms. The 2 additional vector fields accompanying the metric contribution arise in this case from the one-form AdS$_2$ zero-modes that KK reduction of the 5d metric on a (possibly rotating) $S^3/\mathbb{Z}_k$ (with 2-dimensional isometry $U(1)\times U(1)$) yields. Finally, the number of gravitinos is half as many as in the $T^4$ case, because we have half as much supersymmetry.

The macroscopic result \eqref{eq:logcorrection5dHS} is in perfect agreement with our earlier microscopic result in~\eqref{eq:saddleLogMicro}.

\begin{table}[t]
\centering
\begin{tabular}{ |c|c|c|c|c| } 
 \hline
 10d & $g_{\mu\nu}$ & $B_{\mu\nu}$ & $C_{\mu\nu}$ & $C_{\alpha\beta\gamma\delta}^+$\\
 \hline\hline
9d & $g_{\mu\nu}$, $A_\mu$ & $B_{\mu}$, $B_{\mu\nu}$ & $C_{\mu},\ C_{
\mu\nu}$ &$C_{\alpha\beta\gamma}$ or $C_{\alpha\beta\gamma\delta}$ \\
 \hline\hline
5d & $g_{\mu\nu}$, $3A_\mu$ & $3B_{\mu}$, $B_{\mu\nu}$ & $3C_{\mu}, \ C_{\mu\nu}$ &$2C_\alpha,\ 2C_{\alpha\beta}$ \\ 
 \hline
\end{tabular}
\caption{The first row contains  the bosonic fields of IIB SUGRA. The second row is the 9d field content after compactifying on the $S^1$. The 5d field content after further compactification on $\mathrm{HS}$ is in the third row. The additional one-forms in going from 9d to 5d come from the 2 one-cycles of $\mathrm{HS}$, or in the case of $C_\alpha$ from $C_{\alpha\beta\gamma\delta}$ wrapping the 2 three-cycles of $\mathrm{HS}$. The 2 two-forms $C_{\alpha\beta}$ come from $C_{\alpha\beta\gamma\delta}$ wrapping the 2 two-cycles of $\mathrm{HS}$. The Hodge diamond of $\mathrm{HS}$ can be found in Appendix~\ref{app:hodge}.}\label{tab:HS2spec}
\end{table}

\section{Discussion: a status summary of Eberhardt's dualities}\label{sec:Discussion}

In this work we attempted further development of the $\mathcal{N}=(2,2)$ AdS$_3$/CFT$_2$ dualities of~\cite{Eberhardt:2017uup}. We close with a summary of achievements and the remaining challenges we find most pressing.

\subsection{Achievements}

\textbf{On AdS}. As reviewed in the introduction, in all eight cases the bulk BPS KK spectra have been matched with corresponding boundary operators \cite{Eberhardt:2017uup,Datta:2017ert}, and the bulk tree-level and one-loop effective actions on AdS yield the correct boundary central charges at the leading and subleading order respectively \cite{ArabiArdehali:2021iwe,ArabiArdehali:2018mil}.\\

\noindent\textbf{On BTZ}. The bulk tree-level and one-loop effective actions on BTZ yield the leading Bekenstein-Hawking entropy and the subleading logarithmic correction to it. The leading entropy is always microscopically reproduced via general AdS$_3$/CFT$_2$ considerations through Cardy's formula \cite{Strominger:1997eq}. However, among the dualities of Eberhardt only in the $\mathrm{ES}$ case the leading entropy has been microscopically reproduced from a protected supersymmetric index \cite{ArabiArdehali:2021iwe} (in the $\mathrm{ES}$ case the nontrivial supersymmetric index is the elliptic genus). In the $\mathrm{HS}_k$ cases (where the elliptic genus vanishes) we managed in this work to reproduce only a fraction $1-\frac{1}{k}$ of the leading entropy from a helicity-trace index. The subleading logarithmic correction to the macroscopic entropy was reproduced from the index as well in the $\mathrm{ES}$ case \cite{ArabiArdehali:2021iwe}, and here also we found a match at this subleading order.

\subsection{Challenges}

\noindent\textbf{I) The $Q_5$ constraints}. In the ES case, the symmetric orbifold CFT is $\mathrm{sym}^{N}\mathrm{ES}$, with $N=Q_1Q_5+1/2$, where $Q_5$ is the D5 charge coinciding with the number of D5 branes, while $Q_1$ is the D1 charge, related to the number of D1 branes $k_1$ via $Q_1=k_1-Q_5/2$ \cite{ArabiArdehali:2021iwe} (\emph{cf.}~\cite{Green:1996dd}). For the duality to make sense at finite $N$, we should hence have $N=(k_1-Q_5/2)Q_5+1/2\in\mathbb{Z},$ which in turn implies $Q_5$ must be odd. A bulk explanation of the odd $Q_5$ constraint is lacking at the moment. In \cite{ArabiArdehali:2021iwe} it was suggested, in analogy with the $\mathrm{AdS_3}\times \frac{S^3\times T^2}{\mathbb{Z}_2}\times T^2$ case of \cite{Datta:2017ert}, that supersymmetry of the bulk $\alpha'$ corrections might provide such an explanation, but the details remain to be checked. From the point of view of the D1D5 system, it was pointed out in \cite{ArabiArdehali:2021iwe} that odd $Q_5$ guarantees that the number of Coulomb branch moduli of the bound-state brane system agrees with the $\mathcal{N}=(2,2)$ expectation. The question that remains is whether there is a variant of the ES duality for even $Q_5$.

In the HS cases, the symmetric orbifold CFT is $\mathrm{sym}^{N}\mathrm{HS}$, with $N=Q_1Q_5$, where $Q_{1,5}$ are the D1/D5 charges, coinciding with the number of D1/D5 branes \cite{ArabiArdehali:2021iwe}. For the boundary CFT to make sense at finite $N$ therefore no constraint on the charges seems necessary. Whether there are bulk constraints on $Q_5$ in the HS cases, and if yes what is their boundary manifestation, is not completely clear to us. If the considerations of supersymmetry in the $\alpha'$ corrected sector and the number of Coulomb branch moduli of the associated bound-state brane system are indeed relevant, they seem to suggest nontrivial constraints in the HS cases as well. Then the question would be what is their manifestation in the boundary CFT.\\

\noindent\textbf{II) HS microstate counting via indices}. As we saw in Section~\ref{sec:asy degeneracy}, the 1st helicity-trace index of the topologically trivial sector does not capture the full leading entropy in the HS cases. This raises the question whether some other observable (such as defect indices, see \emph{e.g.}~\cite{Chen:2023lzq}) can capture the black hole saddle point. Alternatively, since as mentioned below Eq.~\eqref{eq:entropyB012} our microscopic saddle-point entropy has a nonzero imaginary part which signals large unobstructed bose-fermi cancellations, one may want to consider the topologically non-trivial sectors and turn on the associated fugacities for the momentum and winding charges of the boundary sigma models in order to obstruct those cancellations. We are currently investigating this approach.

Another question is whether a bulk explanation can be given for the factor of $1-\frac{1}{k}$ mismatch with the boundary topologically trivial sector.

It would be interesting to study also the Schwarzian sector of these $\mathcal{N}=(2,2)$ dualities in the context of \cite{Boruch:2022tno}, and in particular see whether the subtle effects arising from non-trivial ratio (denoted $r$ in \cite{Boruch:2022tno}) of the minimal $R$-charge to that of the supercharge can be understood in more detail through the $\mathrm{HS}$ dualities, and possibly be even related to the $1-\frac{1}{k}$ mismatch.

\begin{acknowledgments}

We are indebted to N.~Benjamin, A.~Gustavsson, M.~Heydeman, C.~Keller, P.~Longhi, S.~Murthy, C.~Nazaroglu, V.~Reys, M.~Ro\v{c}ek, and G.~Turiaci for helpful discussions. AA is especially grateful to L.~Eberhardt, J.~Jiang, and W.~Zhao for initial collaboration, and to J.T.~Liu for teaching him to fish with~orbifolds. We also thank the anonymous JHEP referee whose generous and meticulous feedback helped us improve our manuscript. This work was supported in part by the NSF grant PHY-2210533 and the Simons Foundation grants 397411 (Simons Collaboration on the Nonperturbative Bootstrap) and 681267 (Simons Investigator Award).

\end{acknowledgments}

\appendix

\section{Complex hyperelliptic surfaces}
\label{app:hodge}

Complex hyperelliptic surfaces $\mathrm{HS}$ of our interest are obtained as quotients $T^4$. The corresponding supersymmetric sigma models have $\mathcal{N}=(2,2)$ supersymmetry  \cite{Eberhardt:2017uup}. In our low-temperature test of the seed partition functions, we made reference to their Hodge diamonds. These are as follows (compare with the $T^4$ diamond):
\begin{equation}
\mathrm{HS}:\ 
\begin{tabular}{ c c c c c }
  &  & 1 & & \\ 
  & 1 &  & 1 & \\  
 0 &  & 2 & & 0\\    
  & 1 &  & 1 & \\
    &  & 1 & &
\end{tabular},\hspace{2cm}
T^4:\ 
\begin{tabular}{ c c c c c }
  &  & 1 & & \\ 
  & 2 &  & 2 & \\  
 1 &  & 4 & & 1\\     
  & 2 &  & 2 & \\
    &  & 1 & &
\end{tabular}.
\end{equation}

These are also called bi-elliptic because
they admit an elliptic fibration over an elliptic curve. We can view them as a finite quotient $(C\times E)/G$ of the product of a torus $C= S^1\times S^1$ and an elliptic curve $E=\mathbb{C}/\Gamma$. The action of $G$ on $C$ is just a translation as written below. The action of $G$ on $E$ is via some representation $G\rightarrow A(E)$. Denoting $e^{2\pi i/3}$ by $\omega,$ these complex surfaces are completely classified as follows \cite{barth}:\\
\begin{equation}
      \begin{tabular}{c|c|c|c}
    Type \quad & $\Gamma$\quad  & $G$\quad & Action of $G$ on $E$\\ 
    $a1$ & arbitrary & $\mathbb{Z}_2$ & $e \rightarrow -e$\\
      $a2$ & arbitrary & $\mathbb{Z}_2\oplus \mathbb{Z}_2 $& $e \rightarrow -e$\\
       &  & & $e \rightarrow e_1$, where $2 e_1\sim0$\\ 
       $b1$ & $\mathbb{Z}\oplus \mathbb{Z} \omega$ & $\mathbb{Z}_3$ & $e \rightarrow \omega e$\\ 
        $b2$ & $\mathbb{Z}\oplus \mathbb{Z} \omega$ & $\mathbb{Z}_3\oplus \mathbb{Z}_3$ & $e \rightarrow \omega e$\\
         &  &  & $e \rightarrow e+e_1,$ where $\omega e_1\sim e_1$\\
       $c1$ & $\mathbb{Z}\oplus \mathbb{Z} i $ & $\mathbb{Z}_4$ & $e \rightarrow i e$\\ 
        $c2$ & $\mathbb{Z}\oplus \mathbb{Z} i $ & $\mathbb{Z}_4\oplus \mathbb{Z}_2$ & $e \rightarrow i e$\\ 
         &  &  & $e \rightarrow  e+e_1,$ where $i e_1\sim e_1$\\ 
         $d$ & $\mathbb{Z}\oplus \mathbb{Z} \omega$ & $\mathbb{Z}_6$ & $e \rightarrow -\omega e$\\ 
      \end{tabular}\label{tab:HStable}
\end{equation}\\

We have only studied the $a1,b1,c1,d$ cases in this paper, to which we have referred to as $\mathrm{HS}_2,\mathrm{HS}_3,\mathrm{HS}_4,\mathrm{HS}_6$ respectively. Computation of the modified index of the other cases is left for future work.

\section{Details of the seed partition function calculations}\label{app:seed}

\subsection{Derivation of $Z[\mathrm{HS}_2]$}

The starting point is the fact that the orbifold partition function is obtained by summing over the two types of boundary conditions (untwisted or twisted) and projecting on $G$-invariant states (see Eq.~(10.83) in \cite{DiFrancesco:1997nk}):
\begin{equation}
    \mathrm{Tr}_{\mathrm{orb}}=\frac{1}{2}\mathrm{Tr}_+(1+G)+\frac{1}{2}\mathrm{Tr}_-(1+G),\label{eq:orbTrZ2}
\end{equation}
where $+$/$-$ denote the sectors with untwisted/twisted boundary condition around the spatial circle, and the $(1+G)/2$ factors project onto $G$-singlets. This is of course a special case of Eq.~\eqref{eq:orbGeneralFormula} where $k=2.$

The first term on the RHS of Eq.~\eqref{eq:orbTrZ2} is half the $T^4$ partition function which can be found explicitly in \cite{Maldacena:1999bp}. This is how the first term on the RHS of Eq.~(\ref{eq:Zhs2}) is obtained.

The second term in (\ref{eq:Zhs2}) is the contribution of the untwisted sector with $G$-insertion. To see this, note that writing $T^4=T^2\times T^2$, the action of $G$ on the first $T^2$, denoted $E$, is the standard $\mathbb{Z}_2$ orbifold action discussed in Sec.~10.4.3 of the yellow book \cite{DiFrancesco:1997nk}. As explained there, with $G$-insertion the standard $\mathbb{Z}_2$ orbifold partition function does not receive contributions from the topologically non-trivial sectors. Each circle of the $T^2$ contributes $|2\eta(\tau)/\theta_2(\tau)|$, as Eq.~(10.78) of \cite{DiFrancesco:1997nk} for the holomorphic partition function implies.\footnote{Note the typo in Eq.~(10.78) of the old version of \cite{DiFrancesco:1997nk}: the factor of 2 should be inside the square root.} The first $T^2$ hence contributes to $Z^{\text{untwisted}}_{\text{w/\
}G\text{-insertion}}$ as
$|2\eta(\tau)/\theta_2(\tau)|^2$. The fermionic partners contribute $|\theta_2(z,\tau)/\eta(\tau)|^2$. For $z=0$ this is implied by the holomorphic result on the second line of Eq.~(10.46) in \cite{DiFrancesco:1997nk}; the generalization to nonzero $z$ is straightforward once the $R$-charges of the four supersymmetric ground states are understood for instance from Section~13.4.1 of \cite{Hori:2003ic}. See also the derivation of \eqref{eq:HS2ferZg} below. Putting the bosonic and fermionic contributions together, we get from the first supersymmetric $T^2$ sigma model a contribution
\begin{equation}
    \left|\frac{2\eta(\tau)}{\theta_2(\tau)}\right|^2\times \left|\frac{\theta_2(z,\tau)}{\eta(\tau)}\right|^2=4\left|\frac{\theta_2(z,\tau)}{\theta_2(\tau)}\right|^2\label{eq:firstT2}
\end{equation}
to $Z^{\text{untwisted}}_{\text{w/\
}G\text{-insertion}}$.

On the second $T^2$, denoted $C$, the action of $G$-insertion only modifies the topological quantum numbers of the compact bosons (see below). These quantum numbers indicate how many times the two bosons of $T^2$ wind as one goes around the two cycles of the worldsheet torus. The perturbative contribution is not affected by the $G$-insertion, and is therefore given simply by  $\left|\frac{\theta_1(z,\tau)}{\eta^3}\right|^2$. Denoting the modified topological sum as $\Theta^{T^2}_{\text{w/\ }G\text{-insertion}}$ we hence obtain
\begin{equation}
\Theta^{T^2}_{\text{w/\ }G\text{-insertion}}\cdot
\left|\frac{\theta_1(z,\tau)}{\eta^3}\right|^2,\label{eq:secondT2}
\end{equation}
as the contribution of the second supersymmetric $T^2$ factor to $Z^{\text{untwisted}}_{\text{w/\
}G\text{-insertion}}$.

The product of (\ref{eq:firstT2}) and (\ref{eq:secondT2}), multiplied by the factor of $1/2$ (prefactor to $\mathrm{Tr}_+$ in Eq.~\eqref{eq:orbTrZ2}), gives the second term on the right-hand side of (\ref{eq:Zhs2}).\\

Let us now discuss $\Theta^{T^2}_{\text{w/\ }G\text{-insertion}}$ more explicitly. To this end, it will be instructive to consider first the orbifold partition function (\emph{i.e.} with the trace as in (\ref{eq:orbTrZ2})) of a single $2\pi R$-periodic compact scalar $\phi$ orbifolded by $G:\phi\mapsto\phi+\pi R$. Of course $G$ simply reduces $R$ to $R/2$. Therefore we already know what the final answer for the partition function will be: simply replace $R$ by $R/2$ in Eq.~(10.61) of the yellow book \cite{DiFrancesco:1997nk} to get
\begin{equation}
\begin{split}
    Z(R/2)&=\frac{1}{2}\left(\frac{R}{\sqrt{2}}Z_{\mathrm{bos}}(\tau)\sum_{m,m'\in\mathbb{Z}}\mathrm{exp}-\frac{\pi R^2|(m/2)\tau-m'/2|^2}{2\mathrm{Im}\tau}\right)\\
    &=\frac{1}{2}\left((m\mathrm{\ even},m'\mathrm{\ even})+(m\mathrm{\ even},m'\mathrm{\ odd})+(m\mathrm{\ odd},m'\mathrm{\ even})+(m\mathrm{\ odd},m'\mathrm{\ odd})\right).\label{eq:Z(R/2)}
\end{split}
\end{equation}
On the second line, guided by Eq.~(\ref{eq:orbTrZ2}) with which we want to make contact, we have broken up the sum over the topological quantum numbers $m,m'\in\mathbb{Z}$ to four separate sums. The first sum, with $m,m'$ both even, gives back $Z(R)$ precisely as expected from (\ref{eq:orbTrZ2}). Similarly, we expect that the second sum gives the partition function with $G$-insertion, the third sum the partition function of the twisted sector without $G$-insertion, and the fourth sum the partition function of the twisted sector with $G$-insertion. To see this more directly, note that $G$-insertion changes the boundary condition along the ``time'' circle and thus should modify $m'$ which counts the winding of $\phi$ along the time circle. In the twisted sector, on the other hand, we have a modification in the ``spatial'' winding which is counted by $m$.

Writing out $Z_{\mathrm{bos}}(\tau)=\frac{1}{\sqrt{\mathrm{Im}\tau}|\eta(\tau)|^2}$, the factor multiplying $\frac{1}{|\eta(\tau)|^2}$ in the first sum on the second line of (\ref{eq:Z(R/2)}) would be what we would call $\Theta^{S^1}$. It explicitly reads
\begin{equation}
    \Theta^{S^1}=\frac{R}{\sqrt{2\mathrm{Im}\tau}}\sum_{\substack{m\in\mathbb{Z}\\ m' \in\mathbb{Z}}}\mathrm{exp}-\frac{\pi R^2|m\tau-m'|^2}{2\mathrm{Im}\tau},
\end{equation}
where we sum over $m,m'\in\mathbb{Z}$ (instead of only even $m,m'$) because we have replaced $m/2\to m,$ $m'/2\to m'$. Then Poisson resummation of the sum over $m'$ yields the familiar sum over the topological sectors
\begin{equation}
    \Theta^{S^1}=\sum_{\substack{m\in\mathbb{Z}\\ e \in\mathbb{Z}}}q^{(e/R+mR/2)^2/2}\bar{q}^{(e/R-mR/2)^2/2}.\label{eq:ThetaS1}
\end{equation}

Analogously, from the second sum on the second line of (\ref{eq:Z(R/2)}) we get
\begin{equation}
    \Theta^{S^1}_{\text{w/\ }G\text{-insertion}}=\frac{R}{\sqrt{2\mathrm{Im}\tau}}\sum_{\substack{m\in\mathbb{Z}\\ m' \text{\ odd}}}\mathrm{exp}-\frac{\pi R^2|m\tau-m'/2|^2}{2\mathrm{Im}\tau}.\label{eq:ThetaS1wG}
\end{equation}
Writing $\sum_{m'\text{\ odd}}=\sum_{m'}-\sum_{m'\text{\ even}}$, Poisson resummation of the $m'$ sum in (\ref{eq:ThetaS1wG}) now yields
\begin{equation}
    \Theta^{S^1}_{\text{w/\ }G\text{-insertion}}=\sum_{m\in\mathbb{Z}}e^{-2\pi R^2 m^2\tau_2/4}\left(\sum_{e\in\mathbb{Z}}e^{-\frac{4\pi}{a}(e^2+\frac{2be}{4\pi i})}-\sum_{\tilde{e}\text{\ odd}}e^{-\frac{\pi}{a}(\tilde{e}^2+\frac{2b\tilde{e}}{2\pi i})}\right),\label{eq:ThetaS1wG2}
\end{equation}
where $\tau=\tau_1+i\tau_2$, $a=R^2/2\tau_2$, and $b=\pi m R^2\tau_1/\tau_2$. A further simplification yields
\begin{equation}
    \Theta^{S^1}_{\text{w/\ }G\text{-insertion}}=\sum_{m\in\mathbb{Z}}\big(\sum_{e\ \mathrm{even}}-\sum_{e\ \mathrm{odd}}\big)\ q^{(e/R+mR/2)^2/2}\bar{q}^{(e/R-mR/2)^2/2}.\label{eq:ThetaS1wGf}
\end{equation}
This is compatible with our expectation that $G$ acts as $+1$ on the states with $e$ even, and as $-1$ on the states with $e$ odd.

Finally, what we are really interested in is $\Theta^{T^2}_{\text{w/\ }G\text{-insertion}}$, where the superscript $T^2$ stands for the second $T^2$ in $T^4$, namely $C$. Since we take $C$ to be a direct product, we have $\tau^{}_C$ as pure imaginary and $\Theta^{T^2}_{\text{w/\ }G\text{-insertion}}=(\Theta^{S^1}_{\text{w/\ }G\text{-insertion}})^2$.\\

The third and fourth terms of (\ref{eq:Zhs2}) are obtained with similar considerations, and explicit expressions for their $\Theta$ functions can be derived as above. In both cases there are no contributions from the topologically non-trivial sectors of the first $T^2$, and the lattice of $m$ topological charges of the second $T^2$ is shifted by $1/2$. Therefore they do not contribute to the $m=0$ sector of our interest.

\subsection{Derivation of $Z[\mathrm{HS}_{3,4,6}]$}

Recall that $\mathrm{HS}=T^4/G$, with $T^4=E\times C$. For the cases where $G=\mathbb{Z}_k$ the action is
\begin{equation}
G\begin{pmatrix} \phi_{12}\\
\phi_{34}
\end{pmatrix}=\begin{pmatrix} e^{2\pi i/k}\,\phi_{12}\\
\phi_{34}+\frac{2\pi R(1+\tau^{}_C)}{k}\\
\end{pmatrix},\label{eq:GactionZ3}
\end{equation}
where $\phi_{12}$ and $\phi_{34}$ are the complex scalars parametrizing $E$ and $C$ respectively.

As indicated by Eq.~\eqref{eq:orbGeneralFormula}, we need to compute
\begin{equation}
    Z^{g^t\text{-twisted}}_{\text{w/\
}g^j\text{-insertion}}[T^4]=Z^{g^t\text{-twisted}}_{\text{w/\
}g^j\text{-insertion}}[E]\times Z^{g^t\text{-twisted}}_{\text{w/\
}g^j\text{-insertion}}[C],\label{eq:orbT4factorized}
\end{equation}
where we have taken advantage of $T^4=E\times C$ to factorize.

The action of $G$ on $C$ is simply a translation, so the computation of 
\begin{equation}
    Z^{g^t\text{-twisted}}_{\text{w/\
}g^j\text{-insertion}}[C],\label{eq:ZtGjC}
\end{equation}
becomes a simple generalization of the $\mathbb{Z}_2$ case: there is a perturbative part $\left|\frac{\theta_1(z,\tau)}{\eta^3}\right|^2$ insensitive to $G$, while the topological part is again the square of that of $S^1,$ with the latter readable from  the $\mathbb{Z}_k$ analog of Eq.~\eqref{eq:Z(R/2)}.

The action of $G$ on $E$ is more non-trivial, and so is the calculation of
\begin{equation}
    Z^{g^t\text{-twisted}}_{\text{w/\
}g^j\text{-insertion}}[E].\label{eq:ZEtwgi}
\end{equation}
This will be discussed next.

\subsubsection{A digression on $T^2/\mathbb{Z}_k$}

The objects \eqref{eq:ZEtwgi} are the same pieces contributing to the orbifold partition function
\begin{equation}
    Z[T^2/\mathbb{Z}_k]=\frac{1}{k}\big(\sum_{j,t=0}^{k-1} Z^{g^t\text{-twisted}}_{\text{w/\
}g^j\text{-insertion}}[T^2]\big),\label{eq:T2orbFormula}
\end{equation}
when $T^2=E$ and the $\mathbb{Z}_k$ action is as in \eqref{tab:HStable}.

Here we some comments on the structure for general $k$, and then specialize to $k=2,3,4,6$. In fact these are the only allowed values of $k$ because the order-$k$ rotation of the orbifold must be an automorphism of a two-dimensional lattice. This requirement further fixes the complex structure of $E$ to be $\tau^{}_E=i$ in the $\mathbb{Z}_4$ case, while for $\mathbb{Z}_3$ and $\mathbb{Z}_6$ we have $\tau^{}_E=e^{2 \pi i /3}$. See \cite{Dulat:2000ae,Dulat:2000xj} for discussions related to this subsection.

The $\mathcal{N}=(2,2)$ supersymmetric sigma models $T^2/\mathbb{Z}_k$ of our interest contain a  complex scalar $\phi^{+}(z)=\phi_1(z)+ i \phi_2(z)$ and  a complex fermion $\psi^{+}(z)=\psi_1(z)+ i \psi_2(z)$. (The complex conjugate fields $\phi^{-}(z)=\phi_1(z)- i \phi_2(z)$ and   $\psi^{-}(z)=\psi_1(z)- i \psi_2(z)$ give rise to the ``anti-particle'' creation modes quantum mechanically, and these will be important below.)  We also have their anti-holomorphic counterparts, which in the discussions below we usually suppress for brevity. The holomorphic fields and their oscillator modes transform under the $\mathbb{Z}_k$ transformations as  
\begin{eqnarray}
    g^{j} \phi^{\pm}=e^{ \frac{\pm2 \pi i\, j}{k}} \phi^{\pm},\quad \alpha_n^{\pm}\rightarrow e^{\pm\frac{2 \pi i \, j}{k}} \alpha_n^{\pm}, \quad j=1,2,\cdots, k-1,\\
    g^{j} \psi^{\pm} =e^{ \frac{\pm2 \pi i \, j}{k}} \psi^{\pm}, \quad d_n^{\pm}\rightarrow e^{ \pm\frac{2 \pi i \, j}{k}} d_n^{\pm},\quad   j=1,2,\cdots, k-1.
\end{eqnarray}

Let $\mathcal{H}$ be the Hilbert space of the orbifold theory. It is obtained by projecting onto $G$ invariants, via $P=\frac{1}{k} \sum_{j=0}^{k-1}g^j$, the untwisted and twisted sectors of the un-orbifolded theory: $\mathcal{H}=\mathcal{H}_u\bigoplus \mathcal{H}_t$. The piece of $P$ containing $g^j$ will be implemented via $g^j$-twisted boundary conditions along the worldsheet time circle.\\
 
 \textbf{The untwisted sector.} In the untwisted sector, bosonic and fermionic fields satisfy the following boundary conditions (in our convention the worldsheet torus has periodicity 1 along the space direction and $\tau$ along the time direction)
\begin{eqnarray}
    \phi^{+} (1)=\phi^{+}(0)+ \Gamma,\nonumber\\
     \phi^+ (\tau)=g^j \phi^+(0)+  \Gamma,\nonumber
\end{eqnarray}
where $\Gamma$ stands for the target-space lattice generated by $1$ and $\tau^{}_{E}$, while $g^j$ is a generator of $G$. 
The periodicity of the fermions in the Ramond ($+$) and NS ($-$) sectors are
\begin{eqnarray}
    \psi^+(1)=\pm\psi^+ (0),\nonumber\\
     \psi^+(\tau)=\pm g^j\psi^+ (0).\nonumber
\end{eqnarray}
Under these boundary conditions, the solutions of the field equations have the expansions 
\begin{eqnarray}
    \phi^+(z)&=&q^+ -i p^+ \log z+ i \sum_{n \neq 0} \frac{1}{n} \alpha_n^+ z^{-n} ,\\
    \psi^+(z)&=&  \sum_{n} d_n^+ z^{-n}. \,\,\,
\end{eqnarray}
In the bosonic sum $ n\in \mathbb{Z}$, while in the fermionic sum $ n\in \mathbb{Z}$  for the R sector and  $ n\in \mathbb{Z}+1/2$ for the NS sector.\\

\textbf{The twisted sector.} In the sector twisted by $g^t$, with $t=1,\dots,k-1,$ the fields satisfy
\begin{eqnarray}
    \phi^+ (1)=g^t \phi^+(0)+ \Gamma,\nonumber\\
     \phi^+(\tau)=g^j \phi^+(0)+ \Gamma.\nonumber
\end{eqnarray}
For fermions in the R and NS  sector:
\begin{eqnarray}
    \psi^+(1)=\pm g^t\psi^+ (0),\nonumber\\
     \psi^+(\tau)=\pm g^j\psi (0),\nonumber
\end{eqnarray}
with the $+$ sign for the R sector and the $-$ sign for the NS sector.
Under these boundary conditions, the fields have expansions
\begin{eqnarray}
  &&  \phi^+(z)=q_f^+ + i \sum_{n \in Z+t/k} \frac{1}{n} \alpha_n^+ z^{-n},\quad t=1,2,...,k-1,\label{eq:phi+modes} \\
 &&   \psi^+(z)= \sum_{n\in Z+t/k+1/2-s/2} d_n^+ z^{-n} , \,\,\, s=0 \,\,\, \text{for}\,\,  NS,\quad s=1 \,\,\, \text{for}\,\,  R.\label{eq:psi+modes}
\end{eqnarray}
Here $q_f^+$ corresponds to the fixed point values of the complex scalar under the target space quotient.

\subsection*{$T^2/\mathbb{Z}_2$ case}
The action of the primitive group element $g\in \mathbb{Z}_2$ on the bosonic Hilbert space is
\begin{eqnarray}
    g|m_1,m_2,n_1,n_2\rangle = |-m_1,-m_2,-n_1,-n_2 \rangle.
\end{eqnarray}
Here $m_{1,2},n_{1,2}$ are the momentum and winding modes of the complex scalar. The untwisted bosonic Hilbert space $\mathcal{H}_u$ decomposes into subspaces with eigenvalues $\pm 1$ under $g$, schematically as $\mathcal{H}_u=\mathcal{H}_u^+\bigoplus \mathcal{H}_u^-$. These subspaces are spanned by 
\begin{eqnarray}
    \mathcal{H}_u^+&=&\lbrace \alpha^+_{-k_1}\cdots \alpha^+ _{-k_l} \bar{\alpha}^+_{-k_{l+1}}\cdots\bar{\alpha}^+_{-k_{2j}}(1+g)|m_1,m_2,n_1,n_2\rangle\rbrace\nonumber\\
   & +&\lbrace \alpha^+_{-k_1}\cdots \alpha^+ _{-k_l} \bar{\alpha}^+_{-k_{l+1}}\cdots\bar{\alpha}^+_{-k_{2j+1}}(1-g)|m_1,m_2,n_1,n_2\rangle\rbrace\nonumber\\
 \mathcal{H}_u^-&=&\lbrace \alpha^+_{-k_1}\cdots \alpha^+ _{-k_l} \bar{\alpha}^+_{-k_{l+1}}\cdots\bar{\alpha}^+_{-k_{2j+1}}(1+g)|m_1,m_2,n_1,n_2\rangle\rbrace\nonumber\\
   & +&\lbrace \alpha^+_{-k_1}\cdots \alpha^+ _{-k_l} \bar{\alpha}^+_{-k_{l+1}}\cdots\bar{\alpha}^+_{-k_{2j}}(1-g)|m_1,m_2,n_1,n_2\rangle\rbrace\nonumber\\
\end{eqnarray}
where $k_i\in \mathbb{N}$. Above we have suppressed the $\alpha^-,\bar{\alpha}^-$ modes, as they are incorporated similarly.

Various partition functions \eqref{eq:ZEtwgi} can now be computed via elementary 2d CFT methods, with the $t=j=0$ case simply the usual $\mathcal{N}=(2,2)$ $T^2$ partition function.

In the $t=0,j=1$ case the bosonic contributions come from the topological sector $m_1=m_2=n_1=n_2=0$ only. Hence, we need to evaluate the trace with a bunch of creation operators acting on $|0,0,0,0\rangle$. The corresponding multi $\alpha^+_{-n}$ oscillator sum will be of the form $1+e^{i\pi} q^n +(e^{i\pi})^2 q^{2n}+...=\frac{1}{1-e^{i\pi} q^n}$. Thus the multi $\alpha^+_{-n},\bar{\alpha}^+_{-n}$ contribution to the bosonic piece of $Z^{\text{untwisted}}_{\text{w/\
}g\text{-insertion}}[E]$ is
     \begin{eqnarray}
         \prod_{n=1}^{\infty} \frac{1}{1- e^{i\pi} q^n}  \frac{1}{1- e^{-i\pi}\bar{q}^n}\,. 
     \end{eqnarray}

Since the modes $\alpha^-_{-n},\bar{\alpha}^-_{-n}$ transform in a manner conjugate to $\alpha^+_{-n},\bar{\alpha}^+_{-n}$, the multi $\alpha^-_{-n},\bar{\alpha}^-_{-n}$ contribution can be easily added to get the full bosonic piece of $Z^{\text{untwisted}}_{\text{w/\
}g\text{-insertion}}[E]$ as
\begin{eqnarray}
    \frac{(q \bar{q})^{\frac{-1}{12}}}{\prod_{n=1}^\infty (1- e^{ i\pi  } q^n)(1- e^{- i\pi } q^n)(1-e^{-i\pi } \bar{q}^n)(1-e^{ i\pi  } \bar{q}^n)}=4\left|\frac{\eta(\tau)}{\theta_2(\tau)}\right|^2,\label{eq:fullBoseUG}
\end{eqnarray}
where we have also included the Casimir energy contribution $q\bar{q}^{-1/12},$ and on the RHS used the special functions of Appendix~\ref{app:thetas}

As for the contribution of the Dirac fermion, recall that we are interested in the RR sector with a $(-1)^F$ insertion. In the untwisted sector there are zero modes that lead to vacuum degeneracy, as well as nonzero modes that can be filled or unfilled. Their contribution to $Z^{\text{untwisted}}_{\text{w/\
}g\text{-insertion}}[E]$ can be read from Eq.~(11.170) in \cite{Hori:2003ic} as\footnote{Note that $b_\text{there}=z+\frac{j}{k}$, while $a^{}_\text{there}=\frac{t}{k}$. Also we take $(\tilde{a}_\text{there},\tilde{b}_\text{there})=(a_\text{there},b_\text{there}).$}
 \begin{equation}
 \begin{split}
    &(q \bar{q})^{\frac{+1}{12}}\Big(y^{\frac{1}{2}}-e^{- i\pi }y^{\frac{-1}{2}}\Big)\Big(\bar{y}^{\frac{1}{2}}-e^{ i\pi }\bar{y}^{\frac{-1}{2}}\Big)\prod_{n=1}^{\infty} (1-  e^{i \pi }y q^n)(1-  e^{ i\pi }y^{-1}q^n)(1-  e^{ i\pi }\bar{y} \bar{q}^n)(1-  e^{i \pi }\bar{y}^{-1}\bar{q}^n)\\
    &=\left|\frac{\theta_2(z,\tau)}{\eta(\tau)}\right|^2.\label{eq:HS2ferZg}
    \end{split}
 \end{equation}
 
Combining the bosonic and fermionic contributions we get:
\begin{equation}
    Z^{\text{untwisted}}_{\text{w/\
}g\text{-insertion}}[E]=4\left|\frac{\theta_2(z,\tau)}{\theta_2(\tau)}\right|^2.
\end{equation}
 
In the twisted sector, since under $g: \phi^+\rightarrow - \phi^+$ four points of the torus remain fixed, we have 4 twisted ground states coming from the bosonic sector. These have $h=\frac{1}{24}, \bar{h}=\frac{1}{24}$. For general $t,k$ the formula is (\emph{cf.}~\cite{Dulat:2000xj,Dixon:1987bg})
\begin{eqnarray}
     h=\bar{h}=-\frac{1}{12}+\frac{1}{2} \frac{t}{k}\left(1-\frac{t}{k}\right).
\end{eqnarray}
Analogously to \eqref{eq:fullBoseUG} the full bosonic piece of $Z^{g\text{-twisted}}_{\text{w/o\
}g\text{-insertion}}[E]$ is
\begin{eqnarray}
    4\times\frac{(q \bar{q})^{\frac{1}{24}}}{\prod_{n=1}^\infty (1-  q^{n-\frac{1}{2}})(1-  q^{n-\frac{1}{2}})(1- \bar{q}^{n-\frac{1}{2}})(1- \bar{q}^{n-\frac{1}{2}})}=4\left|\frac{\eta(\tau)}{\theta_4(\tau)}\right|^2,\label{eq:fullBoseT}
\end{eqnarray}
with the mode indices now taking half-integer values due to the spatial twist. Similarly, the full bosonic piece of $Z^{g\text{-twisted}}_{\text{w/
}g\text{-insertion}}[E]$ is
\begin{eqnarray}
    4\times\frac{(q \bar{q})^{\frac{1}{24}}}{\prod_{n=1}^\infty (1- e^{ i\pi  } q^{n-\frac{1}{2}})(1- e^{- i\pi } q^{n-\frac{1}{2}})(1-e^{-i\pi } \bar{q}^{n-\frac{1}{2}})(1-e^{ i\pi  } \bar{q}^{n-\frac{1}{2}})}=4\left|\frac{\eta(\tau)}{\theta_3(\tau)}\right|^2.\label{eq:fullBoseTG}
\end{eqnarray}
 
The fermionic sector has the opposite ground state dimensions (\emph{cf.}~Eq.~(11.170) in \cite{Hori:2003ic})
 \begin{eqnarray}
     h=\bar{h}=\frac{1}{12}-\frac{1}{2} \frac{t}{k}\left(1-\frac{t}{k}\right),\label{eq:twCas}
 \end{eqnarray}
and the zero-point charges (\emph{cf.}~Eq.~(11.170) in \cite{Hori:2003ic})
\begin{equation}
    2J_0=2\bar{J}_0=\frac{1}{2}-\frac{t}{k}.\label{eq:twZpCharge}
\end{equation}
Analogously to \eqref{eq:HS2ferZg} the fermionic contribution to $Z^{g\text{-twisted}}_{\text{w/o\
}g\text{-insertion}}[E]$ can be read from Eq.~(11.170) in \cite{Hori:2003ic} as
 \begin{equation}
    (q \bar{q})^{\frac{-1}{24}}\prod_{n=1}^{\infty} (1-  y q^{n-\frac{1}{2}})(1-  y^{-1}q^{n-\frac{1}{2}})(1-  \bar{y} \bar{q}^{n-\frac{1}{2}})(1-  \bar{y}^{-1}\bar{q}^{n-\frac{1}{2}})=\left|\frac{\theta_4(z,\tau)}{\eta(\tau)}\right|^2,\label{eq:HS2ferZt}
 \end{equation}
and similarly the fermionic contribution to $Z^{g\text{-twisted}}_{\text{w/
}g\text{-insertion}}[E]$ is
 \begin{equation}
    (q \bar{q})^{\frac{-1}{24}}\prod_{n=1}^{\infty} (1-  e^{i \pi }y q^{n-\frac{1}{2}})(1-  e^{ i\pi }y^{-1}q^{n-\frac{1}{2}})(1-  e^{ i\pi }\bar{y} \bar{q}^{n-\frac{1}{2}})(1-  e^{i \pi }\bar{y}^{-1}\bar{q}^{n-\frac{1}{2}})=\left|\frac{\theta_3(z,\tau)}{\eta(\tau)}\right|^2.\label{eq:HS2ferZt}
 \end{equation}

Combining the bosonic and fermionic contributions we get:
\begin{equation}
    Z^{g\text{-twisted}}_{\text{w/o\
}g\text{-insertion}}[E]=4\left|\frac{\theta_4(z,\tau)}{\theta_4(\tau)}\right|^2,
\end{equation}
and
\begin{equation}
    Z^{g\text{-twisted}}_{\text{w/
}g\text{-insertion}}[E]=4\left|\frac{\theta_3(z,\tau)}{\theta_3(\tau)}\right|^2.
\end{equation}
 
Although not relevant for our purposes in this work, we can now find $Z[T^2/\mathbb{Z}_2]$ from the formula \eqref{eq:T2orbFormula} as
\begin{equation}
\begin{split}
Z[T^2/\mathbb{Z}_2]&=\frac{1}{2}\Theta^{T^2}\left|\frac{\theta_1(z,\tau)}{\eta^3}\right|^2
+2\left|\frac{\theta_2(z,\tau)}{\theta_2(\tau)}\right|^2
 +2\left|\frac{\theta_4(z,\tau)}{\theta_4(\tau)}\right|^2
 +2\left|\frac{\theta_3(z,\tau)}{\theta_3(\tau)}\right|^2.
\end{split}
\end{equation}
 
\subsection*{$T^2/\mathbb{Z}_3$ case }
In the $\mathbb{Z}_3 $ case the action of the primitive element $g\in \mathbb{Z}_3$ on the Bosonic Hilbert space is 
\begin{eqnarray}
    g|m_1,m_2,n_1,n_2\rangle=|m_2,-m_1-m_2,n_2-n_1,-n_1 \rangle
\end{eqnarray}

The partition functions with $g$ and $g^2$ inserted in the trace will get contributions from the $m_1=m_2=n_1=n_2=0$ sector only. Hence, we need to evaluate the trace with a bunch of creation operators acting on $|0,0,0,0\rangle$. Analogously to \eqref{eq:fullBoseUG} we have the bosonic piece of $Z^{\text{untwisted}}_{\text{w/\
}g^j\text{-insertion}}[E]$ as
 \begin{equation}
    \frac{(q \bar{q})^{-1/12}}{\prod_{n=1} (1- e^{2 \pi i j /3} q^n)(1- e^{-2 \pi i j/3} q^n)(1-e^{-2 \pi i j/3} \bar{q}^n)(1-e^{2 \pi i j/3} \bar{q}^n)}=4\sin^2\frac{\pi j}{3}\left|\frac{\eta(\tau)}{\theta_1(\frac{j}{3},\tau)}\right|^2.
\end{equation}
The fermionic contribution to $Z^{\text{untwisted}}_{\text{w/\
}g^j\text{-insertion}}[E]$ can be written as
\begin{equation}
\begin{split}
     &\Big(y^{\frac{1}{2}}-e^{-\frac{2}{3} \pi i j}y^{\frac{-1}{2}}\Big)\Big(\bar{y}^{\frac{1}{2}}-e^{\frac{2}{3} \pi i j}\bar{y}^{\frac{-1}{2}}\Big)\prod_{n=1}^{\infty} \Big((1-  e^{\frac{2}{3} \pi i j}y q^n)(1-  e^{\frac{-2}{3} \pi i j}y^{-1}q^n)\Big)\\
     &\qquad\qquad\times \Big((1-  e^{\frac{-2}{3} \pi i j}\bar{y} \bar{q}^n)(1-  e^{\frac{2}{3} \pi i j}\bar{y}^{-1}\bar{q}^n)\Big)=\left|\frac{\theta_1(z+\frac{j}{3},\tau)}{\eta(\tau)}\right|^2.
     \end{split}
 \end{equation}
 
Combining the bosonic and fermionic contributions we get:
\begin{equation}
    Z^{\text{untwisted}}_{\text{w/
}g^j\text{-insertion}}[E]=4\sin^2\frac{\pi j}{3}\left|\frac{\theta_1(z+\frac{j}{3},\tau)}{\theta_1(\frac{j}{3},\tau)}\right|^2,\qquad j=1,2.\label{eq:ZHS3UGj}
\end{equation}
 
Let us now consider the twisted sector. For $T^2/\mathbb{Z}_3$, the target space complex structure is $\tau_E=e^{2 \pi i /3}$. There are three fixed points for this orbifold (one at the origin and the other two at the $1/3$ points on the longer diagonal). They lead to triple degeneracy of the bosonic vacua. The conformal weight for the bosonic ground states is $(\frac{1}{36},\frac{1}{36})$. Analogously to \eqref{eq:fullBoseT} the full bosonic piece of $Z^{g^t\text{-twisted}}_{\text{w\
}g^j\text{-insertion}}[E]$ is
\begin{equation}
\begin{split}
    &3\times\frac{(q \bar{q})^{\frac{1}{36}}}{\prod_{n=1}^\infty (1-  e^{-\frac{2\pi i}{k}j}q^{n-\frac{1}{3}})(1-  e^{\frac{2\pi i}{k}j}q^{n-\frac{2}{3}})(1- e^{\frac{2\pi i}{k}j}\bar{q}^{n-\frac{1}{3}})(1- e^{-\frac{2\pi i}{k}j}\bar{q}^{n-\frac{2}{3}})},\quad t=1,\\
    &3\times\frac{(q \bar{q})^{\frac{1}{36}}}{\prod_{n=1}^\infty (1-  e^{-\frac{2\pi i}{k}j}q^{n-\frac{2}{3}})(1-  e^{\frac{2\pi i}{k}j}q^{n-\frac{1}{3}})(1- e^{\frac{2\pi i}{k}j}\bar{q}^{n-\frac{2}{3}})(1- e^{-\frac{2\pi i}{k}j}\bar{q}^{n-\frac{1}{3}})},\quad t=2.
    \label{eq:fullBoseTZ3}
    \end{split}
\end{equation}
In the $t=1$ sector, the factors with $q^{n-2/3}$ come from the creation modes $\alpha^+_{-n_+}$ with $n_+\in\mathbb{Z}_{\ge0}+1/3$ as indicated in \eqref{eq:phi+modes}, while the factors with $q^{n-1/3}$ come from the (target-space complex-) conjugate creation modes $\alpha^-_{-n_-}$ with $n_-\in\mathbb{Z}_{\ge1}-1/3$. In the $t=2$ sector, it is the opposite. The anti-holomorphic factors come from the anti-particle creation modes $\bar{\alpha}^\pm_{-n_\pm}$.

Analogously to \eqref{eq:HS2ferZt} the fermionic contribution to $Z^{g^t\text{-twisted}}_{\text{w\
}g^j\text{-insertion}}[E]$ can be read from Eq.~(11.170) in \cite{Hori:2003ic} as
 \begin{equation}
 \begin{split}
    &(q \bar{q})^{\frac{-1}{36}}(y\bar y)^{\frac{1}{6}}\prod_{n=1}^\infty (1-e^{-\frac{2\pi i}{k}j}  y^{-1}q^{n-\frac{1}{3}})(1-e^{\frac{2\pi i}{k}j}  yq^{n-\frac{2}{3}})(1- e^{\frac{2\pi i}{k}j}\bar y^{-1}\bar{q}^{n-\frac{1}{3}})(1- e^{-\frac{2\pi i}{k}j}\bar y\bar{q}^{n-\frac{2}{3}}),\quad t=1,\\
    &(q \bar{q})^{\frac{-1}{36}}(y\bar y)^{-\frac{1}{6}}\prod_{n=1}^\infty (1-  e^{-\frac{2\pi i}{k}j}y^{-1}q^{n-\frac{2}{3}})(1-e^{\frac{2\pi i}{k}j}  yq^{n-\frac{1}{3}})(1-e^{\frac{2\pi i}{k}j} \bar y^{-1}\bar{q}^{n-\frac{2}{3}})(1- e^{-\frac{2\pi i}{k}j}\bar y\bar{q}^{n-\frac{1}{3}}),\quad t=2.
    \label{eq:HS3ferZt}
     \end{split}
 \end{equation}
In the $t=1$ sector, the factors with $q^{n-2/3}$ come from the creation modes $d^+_{-n_+}$ with $n_+\in\mathbb{Z}_{\ge0}+1/3$ as indicated in \eqref{eq:psi+modes}, while the factors with $q^{n-1/3}$ come from the (target-space complex-) conjugate creation modes $d^-_{-n_-}$ with $n_-\in\mathbb{Z}_{\ge1}-1/3$. In the $t=2$ sector, it is the opposite. The anti-holomorphic factors come from the anti-particle creation modes $\bar{d}^\pm_{-n_\pm}$.

Combining the bosonic and fermionic contributions we get:
\begin{equation}
    Z^{g^t\text{-twisted}}_{\text{w/
}g^j\text{-insertion}}[E]=3(y\bar y)^{1-\frac{t}{3}}\left|\frac{\theta_1(z+\frac{t}{3}\tau+\frac{j}{3},\tau)}{\theta_1(\frac{t}{3}\tau+\frac{j}{3},\tau)}\right|^2,\qquad t=1,2,\ j=0,1,2.\label{eq:ZHS3TGj}
\end{equation}

\subsection*{$T^2/\mathbb{Z}_4$ orbifold}
For the $\mathbb{Z}_4$ orbifold the complex structure of the target space torus is $\tau_E=i$. As before, the bosonic Hilbert space is made out of momentum and winding sectors. The action of the orbifold on these ground states is
\begin{eqnarray}
    g|m_1,m_2,n_1,n_2\rangle =|m_2,-m_1,n_2,-n_1\rangle
\end{eqnarray}

In the untwisted sector we have analogously to \eqref{eq:ZHS3UGj}:
\begin{equation}
    Z^{\text{untwisted}}_{\text{w/
}g^j\text{-insertion}}[E]=4\sin^2\frac{\pi j}{4}\left|\frac{\theta_1(z+\frac{j}{4},\tau)}{\theta_1(\frac{j}{4},\tau)}\right|^2,\qquad j=1,2,3.\label{eq:ZHS4UGj}
\end{equation}

For the twisted sector, we need to find the fixed points. The square lattice has two fixed points under the primitive $g\in\mathbb{Z}_4$, and two further under $g^2$, but the latter two are identified under $g$. Therefore we have two vacua in the $t=1,3$ sectors and three vacua for $t=2$. Analogously to \eqref{eq:fullBoseT} the full bosonic piece of $Z^{g^t\text{-twisted}}_{\text{w\
}g^j\text{-insertion}}[E]$ is
\begin{equation}
\begin{split}
    &2\times\frac{(q \bar{q})^{\frac{1}{96}}}{\prod_{n=1}^\infty (1-  e^{-\frac{2\pi i}{k}j}q^{n-\frac{1}{4}})(1-  e^{\frac{2\pi i}{k}j}q^{n-\frac{3}{4}})(1- e^{\frac{2\pi i}{k}j}\bar{q}^{n-\frac{1}{4}})(1- e^{-\frac{2\pi i}{k}j}\bar{q}^{n-\frac{3}{4}})},\quad t=1,\\
    &3\times\frac{(q \bar{q})^{\frac{1}{24}}}{\prod_{n=1}^\infty (1-  e^{-\frac{2\pi i}{k}j}q^{n-\frac{1}{2}})(1-  e^{\frac{2\pi i}{k}j}q^{n-\frac{1}{2}})(1- e^{\frac{2\pi i}{k}j}\bar{q}^{n-\frac{1}{2}})(1- e^{-\frac{2\pi i}{k}j}\bar{q}^{n-\frac{1}{2}})},\quad t=2,\\
    &2\times\frac{(q \bar{q})^{\frac{1}{96}}}{\prod_{n=1}^\infty (1-  e^{-\frac{2\pi i}{k}j}q^{n-\frac{3}{4}})(1-  e^{\frac{2\pi i}{k}j}q^{n-\frac{1}{4}})(1- e^{\frac{2\pi i}{k}j}\bar{q}^{n-\frac{3}{4}})(1- e^{-\frac{2\pi i}{k}j}\bar{q}^{n-\frac{1}{4}})},\quad t=3.
    \label{eq:fullBoseTZ4}
    \end{split}
\end{equation}
In the $t=1$ sector, the factors with $q^{n-3/4}$ come from the creation modes $\alpha^+_{-n_+}$ with $n_+\in\mathbb{Z}_{\ge0}+1/4$ as indicated in \eqref{eq:phi+modes}, while the factors with $q^{n-1/4}$ come from the (target-space complex-) conjugate creation modes $\alpha^-_{-n_-}$ with $n_-\in\mathbb{Z}_{\ge1}-1/4$. The anti-holomorphic factors come from the anti-particle creation modes $\bar{\alpha}^\pm_{-n_\pm}$. In the $t=3$ sector it is the opposite. The $t=2$ sector is analogous.
Analogously to \eqref{eq:HS2ferZt} the fermionic contribution to $Z^{g^t\text{-twisted}}_{\text{w\
}g^j\text{-insertion}}[E]$ can be read from Eq.~(11.170) in \cite{Hori:2003ic} as
 \begin{equation}
 \begin{split}
    &(q \bar{q})^{\frac{-1}{96}}(y\bar y)^{\frac{1}{4}}\prod_{n=1}^\infty (1-e^{-\frac{2\pi i}{k}j}  y^{-1}q^{n-\frac{1}{4}})(1-e^{\frac{2\pi i}{k}j}  yq^{n-\frac{3}{4}})(1- e^{\frac{2\pi i}{k}j}\bar y^{-1}\bar{q}^{n-\frac{1}{4}})(1- e^{-\frac{2\pi i}{k}j}\bar y\bar{q}^{n-\frac{3}{4}}),\quad t=1,\\
     &(q \bar{q})^{\frac{-1}{24}}\prod_{n=1}^\infty (1-e^{-\frac{2\pi i}{k}j}  y^{-1}q^{n-\frac{1}{2}})(1-e^{\frac{2\pi i}{k}j}  yq^{n-\frac{1}{2}})(1- e^{\frac{2\pi i}{k}j}\bar y^{-1}\bar{q}^{n-\frac{1}{2}})(1- e^{-\frac{2\pi i}{k}j}\bar y\bar{q}^{n-\frac{1}{2}}),\quad t=2,\\
    &(q \bar{q})^{\frac{-1}{96}}(y\bar y)^{-\frac{1}{4}}\prod_{n=1}^\infty (1-  e^{-\frac{2\pi i}{k}j}y^{-1}q^{n-\frac{3}{4}})(1-e^{\frac{2\pi i}{k}j}  yq^{n-\frac{1}{4}})(1-e^{\frac{2\pi i}{k}j} \bar y^{-1}\bar{q}^{n-\frac{3}{4}})(1- e^{-\frac{2\pi i}{k}j}\bar y\bar{q}^{n-\frac{1}{4}}),\quad t=3.
    \label{eq:HS4ferZt}
     \end{split}
 \end{equation}

Combining the bosonic and fermionic contributions we get:
\begin{equation}
\begin{split}
    Z^{g^t\text{-twisted}}_{\text{w/
}g^j\text{-insertion}}[E]&=2(y\bar y)^{1-\frac{t}{4}}\left|\frac{\theta_1(z+\frac{t}{4}\tau+\frac{j}{4},\tau)}{\theta_1(\frac{t}{4}\tau+\frac{j}{4},\tau)}\right|^2,\qquad t=1,3,\ j=0,1,2,3,\\
Z^{g^t\text{-twisted}}_{\text{w/
}g^j\text{-insertion}}[E]&=3(y\bar y)^{1-\frac{t}{4}}\left|\frac{\theta_1(z+\frac{t}{4}\tau+\frac{j}{4},\tau)}{\theta_1(\frac{t}{4}\tau+\frac{j}{4},\tau)}\right|^2,\qquad t=2,\ j=0,1,2,3.
\label{eq:ZHS4TGj}
\end{split}
\end{equation}


\subsection*{$T^2/\mathbb{Z}_6$ orbifold}
For $\mathbb{Z}_6$ orbifold the complex structure of the target space torus is $\tau_E=e^{2 \pi i /3}$. As before, the bosonic Hilbert space is made out of momentum and winding sectors. The action of the orbifold on these ground states is
\begin{eqnarray}
    g|m_1,m_2,n_1,n_2\rangle =|m_1+m_2,-m_1,n_2,-n_1+n_2\rangle.
\end{eqnarray}

In the untwisted sector we have analogously to \eqref{eq:ZHS3UGj}:
\begin{equation}
    Z^{\text{untwisted}}_{\text{w/
}g^j\text{-insertion}}[E]=4\sin^2\frac{\pi j}{6}\left|\frac{\theta_1(z+\frac{j}{6},\tau)}{\theta_1(\frac{j}{6},\tau)}\right|^2,\qquad j=1,2,3,4,5.\label{eq:ZHS6UGj}
\end{equation}

For the twisted sector, we need to find the fixed points. The hexagonal lattice has one fixed point under the primitive $g\in\mathbb{Z}_6$, and two further under $g^2$ (these are the non-origin fixed points of the $\mathbb{Z}_3$ case), but the latter two are identified under $g$. It also has four fixed points under $g^3,$ three of which (namely the non-origin fixed points of the $\mathbb{Z}_2$ case) are identified under $g$. Therefore we have one vacuum in the $t=1,5$ sectors, and two vacua for $t=2,3,4$.  Analogously to \eqref{eq:fullBoseT} the full bosonic piece of $Z^{g^t\text{-twisted}}_{\text{w\
}g^j\text{-insertion}}[E]$ is
\begin{equation}
\begin{split}
    &1\times\frac{(q \bar{q})^{\frac{-1}{72}}}{\prod_{n=1}^\infty (1-  e^{-\frac{2\pi i}{k}j}q^{n-\frac{1}{6}})(1-  e^{\frac{2\pi i}{k}j}q^{n-\frac{5}{6}})(1- e^{\frac{2\pi i}{k}j}\bar{q}^{n-\frac{1}{6}})(1- e^{-\frac{2\pi i}{k}j}\bar{q}^{n-\frac{5}{6}})},\quad t=1,\\
    &2\times\frac{(q \bar{q})^{\frac{1}{36}}}{\prod_{n=1}^\infty (1-  e^{-\frac{2\pi i}{k}j}q^{n-\frac{2}{6}})(1-  e^{\frac{2\pi i}{k}j}q^{n-\frac{4}{6}})(1- e^{\frac{2\pi i}{k}j}\bar{q}^{n-\frac{2}{6}})(1- e^{-\frac{2\pi i}{k}j}\bar{q}^{n-\frac{4}{6}})},\quad t=2,\\
      &2\times\frac{(q \bar{q})^{\frac{1}{24}}}{\prod_{n=1}^\infty (1-  e^{-\frac{2\pi i}{k}j}q^{n-\frac{3}{6}})(1-  e^{\frac{2\pi i}{k}j}q^{n-\frac{3}{6}})(1- e^{\frac{2\pi i}{k}j}\bar{q}^{n-\frac{3}{6}})(1- e^{-\frac{2\pi i}{k}j}\bar{q}^{n-\frac{3}{6}})},\quad t=3,\\
        &2\times\frac{(q \bar{q})^{\frac{1}{36}}}{\prod_{n=1}^\infty (1-  e^{-\frac{2\pi i}{k}j}q^{n-\frac{4}{6}})(1-  e^{\frac{2\pi i}{k}j}q^{n-\frac{2}{6}})(1- e^{\frac{2\pi i}{k}j}\bar{q}^{n-\frac{4}{6}})(1- e^{-\frac{2\pi i}{k}j}\bar{q}^{n-\frac{2}{6}})},\quad t=4,\\
    &1\times\frac{(q \bar{q})^{\frac{-1}{72}}}{\prod_{n=1}^\infty (1-  e^{-\frac{2\pi i}{k}j}q^{n-\frac{5}{6}})(1-  e^{\frac{2\pi i}{k}j}q^{n-\frac{1}{6}})(1- e^{\frac{2\pi i}{k}j}\bar{q}^{n-\frac{5}{6}})(1- e^{-\frac{2\pi i}{k}j}\bar{q}^{n-\frac{1}{6}})},\quad t=5.
    \label{eq:fullBoseTZ6}
    \end{split}
\end{equation}
Analogously to \eqref{eq:HS2ferZt} the fermionic contribution to $Z^{g^t\text{-twisted}}_{\text{w\
}g^j\text{-insertion}}[E]$ can be read from Eq.~(11.170) in \cite{Hori:2003ic} as
 \begin{equation}
 \begin{split}
    &(q \bar{q})^{\frac{1}{72}}(y\bar y)^{\frac{1}{3}}\prod_{n=1}^\infty (1-e^{-\frac{2\pi i}{k}j}  y^{-1}q^{n-\frac{1}{6}})(1-e^{\frac{2\pi i}{k}j}  yq^{n-\frac{5}{6}})(1- e^{\frac{2\pi i}{k}j}\bar y^{-1}\bar{q}^{n-\frac{1}{6}})(1- e^{-\frac{2\pi i}{k}j}\bar y\bar{q}^{n-\frac{5}{6}}),\quad t=1,\\
     &(q \bar{q})^{\frac{-1}{36}}(y\bar y)^{\frac{1}{6}}\prod_{n=1}^\infty (1-e^{-\frac{2\pi i}{k}j}  y^{-1}q^{n-\frac{2}{6}})(1-e^{\frac{2\pi i}{k}j}  yq^{n-\frac{4}{6}})(1- e^{\frac{2\pi i}{k}j}\bar y^{-1}\bar{q}^{n-\frac{2}{6}})(1- e^{-\frac{2\pi i}{k}j}\bar y\bar{q}^{n-\frac{4}{6}}),\quad t=2,\\
    &(q \bar{q})^{\frac{-1}{24}}\prod_{n=1}^\infty (1-e^{-\frac{2\pi i}{k}j}  y^{-1}q^{n-\frac{3}{6}})(1-e^{\frac{2\pi i}{k}j}  yq^{n-\frac{3}{6}})(1- e^{\frac{2\pi i}{k}j}\bar y^{-1}\bar{q}^{n-\frac{3}{6}})(1- e^{-\frac{2\pi i}{k}j}\bar y\bar{q}^{n-\frac{3}{6}}),\quad t=3,\\
     &(q \bar{q})^{\frac{-1}{36}}(y\bar y)^{-\frac{1}{6}}\prod_{n=1}^\infty (1-e^{-\frac{2\pi i}{k}j}  y^{-1}q^{n-\frac{4}{6}})(1-e^{\frac{2\pi i}{k}j}  yq^{n-\frac{2}{6}})(1- e^{\frac{2\pi i}{k}j}\bar y^{-1}\bar{q}^{n-\frac{4}{6}})(1- e^{-\frac{2\pi i}{k}j}\bar y\bar{q}^{n-\frac{2}{6}}),\quad t=4,\\
      &(q \bar{q})^{\frac{1}{72}}(y\bar y)^{-\frac{1}{3}}\prod_{n=1}^\infty (1-e^{-\frac{2\pi i}{k}j}  y^{-1}q^{n-\frac{5}{6}})(1-e^{\frac{2\pi i}{k}j}  yq^{n-\frac{1}{6}})(1- e^{\frac{2\pi i}{k}j}\bar y^{-1}\bar{q}^{n-\frac{5}{6}})(1- e^{-\frac{2\pi i}{k}j}\bar y\bar{q}^{n-\frac{1}{6}}),\quad t=5.
    \label{eq:HS6ferZt}
     \end{split}
 \end{equation}
 Combining the bosonic and fermionic contributions we get:
\begin{equation}
\begin{split}
    Z^{g^t\text{-twisted}}_{\text{w/
}g^j\text{-insertion}}[E]&=(y\bar y)^{1-\frac{t}{6}}\left|\frac{\theta_1(z+\frac{t}{6}\tau+\frac{j}{6},\tau)}{\theta_1(\frac{t}{6}\tau+\frac{j}{6},\tau)}\right|^2,\qquad t=1,5,\quad j=0,1,2,3,4,5,\\
 Z^{g^t\text{-twisted}}_{\text{w/
}g^j\text{-insertion}}[E]&=2(y\bar y)^{1-\frac{t}{6}}\left|\frac{\theta_1(z+\frac{t}{6}\tau+\frac{j}{6},\tau)}{\theta_1(\frac{t}{6}\tau+\frac{j}{6},\tau)}\right|^2,\qquad t=2,3,4,\quad j=0,1,2,3,4,5.\\
\label{eq:ZHS6TGj}
\end{split}
\end{equation}

\subsubsection{Compiling the pieces}

As implied by \eqref{eq:orbGeneralFormula} and \eqref{eq:orbT4factorized}, we have
\begin{equation}
    Z[\mathrm{HS}_k]=\frac{1}{k}\big(\sum_{j,t=0}^{k-1} Z^{g^t\text{-twisted}}_{\text{w/\
}g^j\text{-insertion}}[E]\times Z^{g^t\text{-twisted}}_{\text{w/\
}g^j\text{-insertion}}[C]\big).
\end{equation}

As explained around \eqref{eq:ZtGjC}, the computation of $Z^{g^t\text{-twisted}}_{\text{w/ }g^j\text{-insertion}}[C]$ is rather straightforward. In the $\mathrm{HS}_2$ case for instance, compiling the above  results for $Z^{g^t\text{-twisted}}_{\text{w/ }g^j\text{-insertion}}[E]$ and $Z^{g^t\text{-twisted}}_{\text{w/ }g^j\text{-insertion}}[C]$ we arrive at
\begin{equation}
\begin{split}
Z[\mathrm{HS}_2]&=\frac{1}{2}\Theta^{T^4}\left|\frac{\theta_1(z,\tau)}{\eta^3}\right|^4
+2\left|\frac{\theta_2(z,\tau)}{\theta_2(\tau)}\right|^2\cdot
\Theta^{T^2}_{\text{w/\ }g\text{-insertion}}\cdot
\left|\frac{\theta_1(z,\tau)}{\eta^3}\right|^2\\
&\ \ +2\left|\frac{\theta_4(z,\tau)}{\theta_4(\tau)}\right|^2\cdot
\Theta^{T^2}_{\text{w/\ }1/2-\text{shifted lattice}}\cdot
\left|\frac{\theta_1(z,\tau)}{\eta^3}\right|^2\\
&\ \ +2\left|\frac{\theta_3(z,\tau)}{\theta_3(\tau)}\right|^2\cdot
\Theta^{T^2}_{\text{w/\ }g\text{-insertion and }1/2-\text{shifted
lattice}}\cdot \left|\frac{\theta_1(z,\tau)}{\eta^3}\right|^2.
\end{split}
\end{equation}

Similarly, for $\mathrm{HS}_{3,4,6}$ we arrive at \eqref{eq:Zhs3}, \eqref{eq:Zhs4}, \eqref{eq:Zhs6} respectively.

\section{Theta functions and their asymptotics}\label{app:thetas}

The Jacobi theta functions are defined as
\begin{equation}
    \begin{split}
        \theta_1(z,\tau)&=-\vartheta_{11}(y,q)=-i\sum_{n\in\mathbb{Z}}(-1)^n y^{n+\frac{1}{2}} q^{(n+\frac{1}{2})^2/2}=-i y^{\frac{1}{2}} q^{\frac{1}{8}} \prod_{n=1}^{\infty} (1-q^n)(1-y q^n)(1- y^{-1} q^{n-1}),\\
        \theta_2(z,\tau)&=\vartheta_{10}(y,q)=\sum_{n\in\mathbb{Z}} y^{n+\frac{1}{2}} q^{(n+\frac{1}{2})^2/2}=y^{\frac{1}{2}} q^{\frac{1}{8}} \prod_{n=1}^{\infty} (1-q^n)(1+y q^n)(1+y^{-1} q^{n-1}),\\
        \theta_3(z,\tau)&=\vartheta_{00}(y,q)=\sum_{n\in\mathbb{Z}} y^{n} q^{n^2/2}= \prod_{n=1}^{\infty} (1-q^n)(1+y q^{n-\frac{1}{2}})(1+ y^{-1} q^{n-\frac{1}{2}}),\\
        \theta_4(z,\tau)&=\vartheta_{01}(y,q)=\sum_{n\in\mathbb{Z}}(-1)^n y^{n} q^{n^2/2}=\prod_{n=1}^{\infty} (1-q^n)(1-y q^{n-\frac{1}{2}})(1- y^{-1} q^{n-\frac{1}{2}}).
    \end{split}
\end{equation}

The transformation formula of $\theta_1$ reads (up to a constant phase)
\begin{equation}
   \theta_1(\frac{u}{\tilde{\tau}};\frac{a\tau+b}{c\tau+d})=\sqrt{-i\tilde{\tau}}e^{i\pi c u^2/\tilde{\tau}}\theta_1(u;\tau),
\end{equation}
where $\tilde{\tau}=c\tau+d$. This, using the fact that  $\theta_1(u,\tau)=iq^{\frac{1}{8}}e^{-i\pi u}(q;q)\theta_0(u,\tau)$ implies
\begin{equation}
\begin{split}
\theta_1(u;\tau)&=\frac{1}{\sqrt{-i\tilde{\tau}}}\,e^{-i\pi c u^2/\tilde{\tau}}e^{\frac{2\pi i}{8}\frac{a\tau+b}{\tilde{\tau}}}e^{-i\pi u/\tilde{\tau}}(\tilde{q};\tilde{q})\,\theta_0(\frac{u}{\tilde{\tau}};\frac{a\tau+b}{c\tau+d}),\label{eq:theta0modularTrans}
\end{split}
\end{equation}
again up to a constant phase. Note that
\begin{equation}
    \theta_0(u;\tau)=\prod_{k=0}^\infty (1-e^{2\pi i(u+k\tau)})(1-e^{2\pi i(-u+(k+1)\tau)}).
\end{equation}
As a result, for $u\in\mathbb{R}$ we have the $\tau\to-d/c$ asymptotics
\begin{equation}
    \theta_1(u;\tau)\approx e^{\frac{i\pi}{c\tilde{\tau}}[\vartheta(cu)-\frac{1}{4}]},
\end{equation}
where $\vartheta(\cdot)=\{\cdot\}(1-\{\cdot\})$, and $\{\cdot\}:=\cdot-\lfloor\cdot\rfloor.$ The asymptotics of $\theta_{2,3,4}$ then follow as
\begin{equation}
    \begin{split}
        \theta_2(u;\tau)&=\theta_1(u+\frac{1}{2};\tau)\approx e^{\frac{i\pi}{c\tilde{\tau}}[\vartheta(cu+\frac{c}{2})-\frac{1}{4}]},\\
        \theta_3(u;\tau)&=y^{-\frac{1}{2}}q^{\frac{1}{8}}\theta_2(u-\frac{\tau}{2};\tau)\approx e^{\frac{i\pi}{c\tilde{\tau}}[\vartheta(cu+\frac{c}{2}+\frac{d}{2})-\frac{1}{4}]},\\
        \theta_4(u;\tau)&=\theta_3(u+\frac{1}{2};\tau)\approx e^{\frac{i\pi}{c\tilde{\tau}}[\vartheta(cu+\frac{d}{2})-\frac{1}{4}]}.
    \end{split}
\end{equation}

The analogous formula for $\eta(\tau):=q^{1/24}\prod_{n=1}^\infty (1-q^n)$ is (see \emph{e.g.} Appendix~A of \cite{ArabiArdehali:2021nsx}):
\begin{equation}
    \eta(\tau)\approx e^{-\frac{i\pi}{12c\tilde{\tau}}}.
\end{equation}

The most important case for us in the main text is of course $d=0,$ $c=1.$

\section{Fine-grained discrete structure in modified indices}\label{app:fine-grained}

For simplicity of exposition we suppressed in the main text a rather intricate---though possibly physically significant---structure in the coefficients $d^{\,\mathrm{sym}N}(n,j)$ of the modified indices 
\begin{equation}
    \mathcal{E}_1[\mathrm{sym}^N\mathrm{HS}_k](q,y)=\sum_{n,j} d^{\,\mathrm{sym}N}(n,j)\ q^n \,y^j.\label{eq:dsymDef}
\end{equation}
For example, in the $\mathrm{HS}_2$ case it turns out (as one can easily check from \eqref{eq:HS2wjf}) that odd powers of $q$ are absent in the counting function $\mathcal{H}_1(q,y)$. This implies via \eqref{eq:E1cHat1} that
\begin{equation}
    \text{$\mathcal{E}_1[\mathrm{sym}^N\mathrm{HS}_2]$ encodes no states with $Nn$ an odd number!}\label{eq:strange1st}
\end{equation}
While in the main text we turned our head away from peculiar ``number theoretic'' phenomena such as this, in the present appendix we explore them systematically.

The fine-grained structure is simplest for $\mathrm{HS}_2$, so we focus on this case. We begin with the Fourier coefficients $\hat{c}_1(n,j)$ of the counting function  $\mathcal{H}_1[\mathrm{HS}_2]$ as found in Section~\ref{subsec:Rademacher}:
\begin{equation}
    \begin{tabular}{|c|c|c|}
    \hline
       $j\backslash n$  & odd & even \\
       \hline
       odd  & $\hat{c}_1=0$ & $\hat{c}_1\approx \pm\exp\big(\pi\sqrt{n-\frac{j^2}{4}}\big)$ \\
        even & $\hat{c}_1=0$&$\hat{c}_1=0$\\ 
        \hline
    \end{tabular}\label{eq:HS2table}
\end{equation}
with the approximation valid as $n\sim j^2\to\infty$.

To go from the coefficients $\hat{c}_1$ just discussed, to our desired coefficients of the (modified) index $\mathcal{E}_1(q,y)$, we use the (modified) DMVV formula \eqref{eq:modifiedDMVV}, together with \eqref{eq:dsymDef}:
\begin{equation}
 \sum_{N=1}^\infty p^N \sum_{n,j} d^{\,\mathrm{sym}N}(n,j)\ q^n \,y^j=\sum_{s,i \in \mathbb{N}, \Delta\geq 0,\ell} \hat{c}_1(i\Delta,\ell)( p^i q^\Delta y^{\ell})^s .\label{eq:modifiedDMVVapp}
\end{equation}

In the main text we limited attention to the $s=1$ terms on the RHS of \eqref{eq:modifiedDMVVapp}. We now explore the contributions from $s$ not necessarily equal to $1$. It follows from \eqref{eq:modifiedDMVVapp} that
\begin{equation}
    d^{\,\mathrm{sym}N}(n,j)=\sum_{s\in\mathrm{Fac}N}d_s^{\,\mathrm{sym}N}(n,j),\label{eq:sDecomposed}
\end{equation}
where $\mathrm{Fac}N=\{1,\dots,N\}$ is the set of natural factors of $N$, and
\begin{equation}
    d_s^{\,\mathrm{sym}N}(n,j)=\hat{c}_1\big(\frac{Nn}{s^2},\frac{j}{s}\big).\label{eq:ds}
\end{equation}

Having \eqref{eq:HS2table}, \eqref{eq:sDecomposed}, and \eqref{eq:ds} we can now state the systematic generalization of \eqref{eq:strange1st}:
\begin{equation}
    \text{in the  $\mathrm{HS}_2$ case } d^{\,\mathrm{sym}N}(n,j)\begin{cases}
        \approx\sum_s\pm\, e^{\frac{\pi}{s}\sqrt{Nn-\frac{j^2}{4}}}\quad\text{for $s\in\mathrm{Fac}N$ s.t. $\frac{Nn}{s^2}\in 2\mathbb{N}$, $\frac{j}{s}\in2\mathbb{Z}+1$,}\\
        =0\quad\text{if no such $s$ exists.}
    \end{cases}\label{eq:generalStrange}
\end{equation}
For example, for $j$ even (of the form $2^m\times\text{odd}$ with $m\in\mathbb{N}$) it follows from \eqref{eq:generalStrange} that
\begin{equation}
     d^{\,\mathrm{sym}N}(n,j)\begin{cases}
        \approx\pm\, e^{\frac{\pi}{2^m}\sqrt{Nn-\frac{j^2}{4}}}\quad\text{if $Nn\in 2^{2m+1}\mathbb{N}$,}\\
        =0\quad\text{otherwise,}
    \end{cases}\label{eq:jStrange}
\end{equation}
where on the first line we have suppressed subleading terms that might arise from $s>2^m$.

\bibliographystyle{JHEP}
\bibliography{refs}

\end{document}